\documentclass[]{aastex631}
\usepackage{comment}
\usepackage{bm}

\begin{document}

\title{Diagnosis of Circumstellar Matter Structure in  Interaction-Powered Supernovae with Hydrogen Line Feature}

\correspondingauthor{Ayako T. Ishii}
\email{ayako-ishii@sci.kj.yamagata-u.ac.jp}

\author[0000-0002-8474-4781]{Ayako T. Ishii}
\affiliation{Faculty of Science, Yamagata University, 1-4-12 Kojirakawa-Machi, Yamagata Yamagata 990-8560, Japan}

\author[0000-0002-8215-5019]{Yuki Takei}
\affiliation{Research Center for the Early Universe (RESCEU), Graduate School of Science, The University of Tokyo, 7-3-1 Hongo, Bunkyo-ku, Tokyo 113-0033, Japan}
\affiliation{Astrophysical Big Bang Laboratory, RIKEN, 2-1 Hirosawa, Wako, Saitama 351-0198, Japan}

\author[0000-0002-6347-3089]{Daichi Tsuna}
\affiliation{TAPIR, Mailcode 350-17, California Institute of Technology, Pasadena, CA 91125, USA}
\affiliation{Research Center for the Early Universe (RESCEU), Graduate School of Science, The University of Tokyo, 7-3-1 Hongo, Bunkyo-ku, Tokyo 113-0033, Japan}

\author[0000-0002-4060-5931]{Toshikazu Shigeyama}
\affiliation{Research Center for the Early Universe (RESCEU), Graduate School of Science, The University of Tokyo, 7-3-1 Hongo, Bunkyo-ku, Tokyo 113-0033, Japan}
\affiliation{Department of Astronomy, School of Science, The University of Tokyo, 7-3-1 Hongo, Bunkyo-ku, Tokyo 113-0033, Japan}

\author[0000-0002-6705-6303]{Koh Takahashi}
\affiliation{National Astronomical Observatory of Japan, National Institutes for Natural Science, 2-21-1 Osawa, Mitaka, Tokyo 181-8588, Japan}








\begin{abstract}

Some supernovae (SNe) are powered by collision of the SN ejecta with a dense circumstellar matter (CSM). Their emission spectra show characteristic line shapes of combined broad emission and narrow P-Cyg lines, which should closely relate to the CSM structure and the mass-loss mechanism that creates the dense CSM. 
We quantitatively investigate the relationship between the line shape and the CSM structure by Monte Carlo radiative transfer simulations, considering two representative cases of dense CSM formed by steady and eruptive mass loss. Comparing the H$\alpha$ emission between the two cases, we find that a narrow P-Cyg line appears in the eruptive case while it does not appear in the steady case, due to the difference in the velocity gradient in the dense CSM. We also reproduce the blue-shifted photon excess observed in some Type IIn SNe, which is formed by photon transport across the shock wave and find the relationship between the velocity of the shocked matter and the amount of the blue shift of the photon excess. 
We conclude that the presence or absence of narrow P-Cyg lines can distinguish the mass loss mechanism, and suggest high-resolution spectroscopic observations with $\lambda/ \Delta \lambda \gtrsim 10^4$ after the light curve peak for applying this diagnostic method. 
\end{abstract}



\section{Introduction} \label{sec_intro}

Supernovae (SNe), which are explosions following the deaths of massive stars, display a diversity in their light curves and spectra that reflects their diverse environment. 
A fraction of SNe exhibit narrow lines in their spectra, and such SNe are thought to be powered by the interaction between dense circumstellar material (CSM) and SN ejecta \citep{Grasberg86,chugai1991,chugai2001,Chevalier11}.
The CSM expands much slower than the SN ejecta, and is believed to produce the narrow lines in the spectra due to the Doppler effect of its line emission.

Estimations from light curves \citep[e.g.,][]{moriya2013,Moriya14} and spectra \citep[e.g.,][]{kiewe2012,taddia2013} of Type IIn SNe (SNe IIn), whose spectra exhibit narrow lines of hydrogen, indicate that their progenitors should have lost mass at a high mass-loss rate of $\dot{M} \gtrsim 10^{-3}\ \rm{M_{\odot}/yr}$ to form the dense CSM.
This is in stark contrast to late time radio observations of SNe with mass-loss rate estimates of $\dot{M} = 10^{-7} - 10^{-5}\ \rm{M_\odot/yr}$
\citep{Weiler02,Chevalier2006,Bietenholz21}, compatible with those observed in stellar winds of local evolved massive stars \citep[e.g.,][]{DeJager88,vanLoon05,Mauron11,Beasor20}. 
Therefore, progenitors of SNe IIn should have undergone a different stage of enhanced mass loss just before the SN explosion\footnote{More recently, short-cadence observations have revealed other types of SNe that are more common than SNe IIn to also have dense CSM; e.g. Type II-P SN like SN 2013fs \citep{yaron2017} and Type Ic SN like SN 2020oi \citep{maeda2021, gagliano2022}}.

One of the plausible formation processes of such dense CSM is the mass ejection from a star prior to the SN explosion.
Existence of such violent mass ejections was pointed out by \cite{chugai2004}, by modeling of H$\alpha$ lines in the SN IIn 1994W using radiative transfer computations. 
Indeed, observations detected precursor outbursts in a significant fraction of SNe IIn \citep{Ofek14,Strotjohann21}, the most famous one being SN 2009ip \citep[e.g.,][]{pastorello2013,Margutti14,Smith14_2009ip}. The outbursts should be strongly linked to the formation of the dense CSM \citep{Ofek14,matsumoto2022}, since the precursor emission ($10^{40}$--$10^{41}$ erg s$^{-1}$) is much brighter than the Eddington limit of a massive star. However, the enhanced mass loss may also be reproduced by steady processes, such as constant near-surface energy deposition \citep{Quataert16}, pulsation \citep[e.g.,][]{Yoon10} or core neutrino emission \citep{Moriya14_neutrino}.
While observations are in great progress, we do not have a clear criterion to distinguish between steady and eruptive mass ejection from the observed features of the CSM. 

A clue to distinguishing these mechanisms may be the structure of the resulting CSM. While the steady mass loss simply predicts a CSM with a density profile close to $\rho(r)\propto r^{-2}$ with constant velocity, the CSM from eruptive mass-loss has recently been investigated in detail. Recent numerical and analytical studies \citep{kuriyama2020, kuriyama2021, Leung20, Linial21, tsuna2021, ko2022, Tsang22,Tsuna_Takei23} have shown that the CSM consists of a bound and unbound part, whose masses strongly depend on the injected energy that triggers the eruption\footnote{For injections with energies much smaller than the binding energy of the envelope, the whole envelope can remain bound, and inflate but not erupt \citep[e.g.,][]{Leung20,ko2022,Wu22}. Such cases would not result in CSM with large radii that are seen in Type IIn SNe, and hence are not the subject of this study.}. After several dynamical timescales of the progenitor, the inner region of the CSM gravitationally bound to the star has a shallower density profile with $\rho(r)\propto r^{-1.5}$, 
and falls back to the star with a velocity of $v(r)\approx -\sqrt{2GM_*/r}$, where $G$ denotes the gravitational constant and $M_*$ is the stellar mass. In contrast to steady-state mass loss with a nearly constant velocity, the CSM created by mass eruption is expected to have a positive velocity gradient, with the outermost region expanding homologously ($v(r)\propto r$).

Such differences of the CSM structure in the two mechanisms may be observable in the line emission. 
Especially, in SNe IIn, the H$\alpha$ lines have a characteristic shape, with a combination of a broad component and a narrow component with a P-Cyg profile \citep{salamanca1998, pastorello2002, kiewe2012,zhang2012, taddia2013, 2014ApJ...797..118F,taddia2020}. Some observed H$\alpha$ lines in the late phase show a blue-shifted excess resulting in an asymmetric shape, which might originate from a region with strong radiative cooling immediately behind the shock front \citep[cool dense shell;][]{dessart2015}, or from the aspherical structures of the CSM \citep{smith2015, kokubo2019}. 

The origin of these spectral components has been investigated in several previous works.
\cite{chugai2001} and \cite{huang2018} have shown by Monte Carlo radiative transfer simulations that the broad component is formed as a result of multiple electron scattering for an optically-thick CSM (or by radiative acceleration at just after shock breakout; \citealt{Kochanek19}), while it is attributed to the bulk motion of the shocked region in the optically thin case. 
However they used a simple model for the shock propagation in the CSM, and also did not include line absorption in their opacity that characterizes the narrow P-Cyg feature. 
By Monte Carlo simulations including the narrow P-Cyg profile, \cite{chugai2004} better reproduced line shapes as seen in the spectrum of SN 1994W assuming homologously expanding CSM rather than CSM formed by steady mass loss, and argued that the progenitor underwent an explosive mass loss event \citep[see also][]{2009MNRAS.394...21D}. 
The simulations of SNe IIn by \cite{dessart2015} have solved equations of radiative transfer by the $\Lambda$-iteration method, taking into account effects of the velocity gradient on the spectral shape of lines (see also \citealt{Groh14,Boian19,boian2020} for applications to early phase spectra). While the simulations generally reproduced the observations of a particular super-luminous Type IIn SN 2010jl, they were limited to CSM of a steady wind-line profile with a constant initial velocity.
The relationship between the line shape and matter structure has also been investigated for Luminous Blue Variables (LBVs) in \citet{groh2011}, which shows that abrupt variation in velocity profile is necessary to reproduce the H$\alpha$ line shape characterising LBVs.

In this study, we quantitatively investigate the relationship between line shapes and CSM structures in order to establish a method to diagnose the CSM structure from the line shapes. 
We investigate the H$\alpha$ line as a first step because it is prominently observed in a large fraction of interaction-powered SNe, including SN 1998S which we focus on as a reference for modeling. 
The propagation of a shock wave in the CSM is calculated by a radiation hydrodynamics code CHIPS \citep{takei2022,Takei_et_al_2023} for both cases of eruptive and steady mass loss, and the spectra are calculated using a Monte Carlo radiative transfer code as post-process. 
Then, the shapes of H$\alpha$ lines in the spectra are compared between eruptive and steady mass-loss cases. We adopt a Monte Carlo approach, which investigates the shape of the H$\alpha$ line that reaches the observer by following the trajectory of each packet of photons \citep[e.g.,][]{lucy2005}. The Monte Carlo approach is appropriate to diagnose the origin of each feature in the spectra, because one can trace back the history of the packets that form the feature.

This paper is organized as follows. 
We present our models for the simulations and the detailed computational methods in Section \ref{sec:model}. 
In Section \ref{sec:result} we present the spectra obtained by our simulations, with detailed discussion on the interplay of the processes of line emission, scattering and absorption. We furthermore compare the line shapes from the CSM formed by eruptive and steady mass-loss scenarios. We discuss our results in the context of observed Type IIn SNe and summarize this paper in Section \ref{sec:conclusion}.

\section{Methods and models} \label{sec:model}

\begin{figure}
    \centering
    \includegraphics[width=150mm]{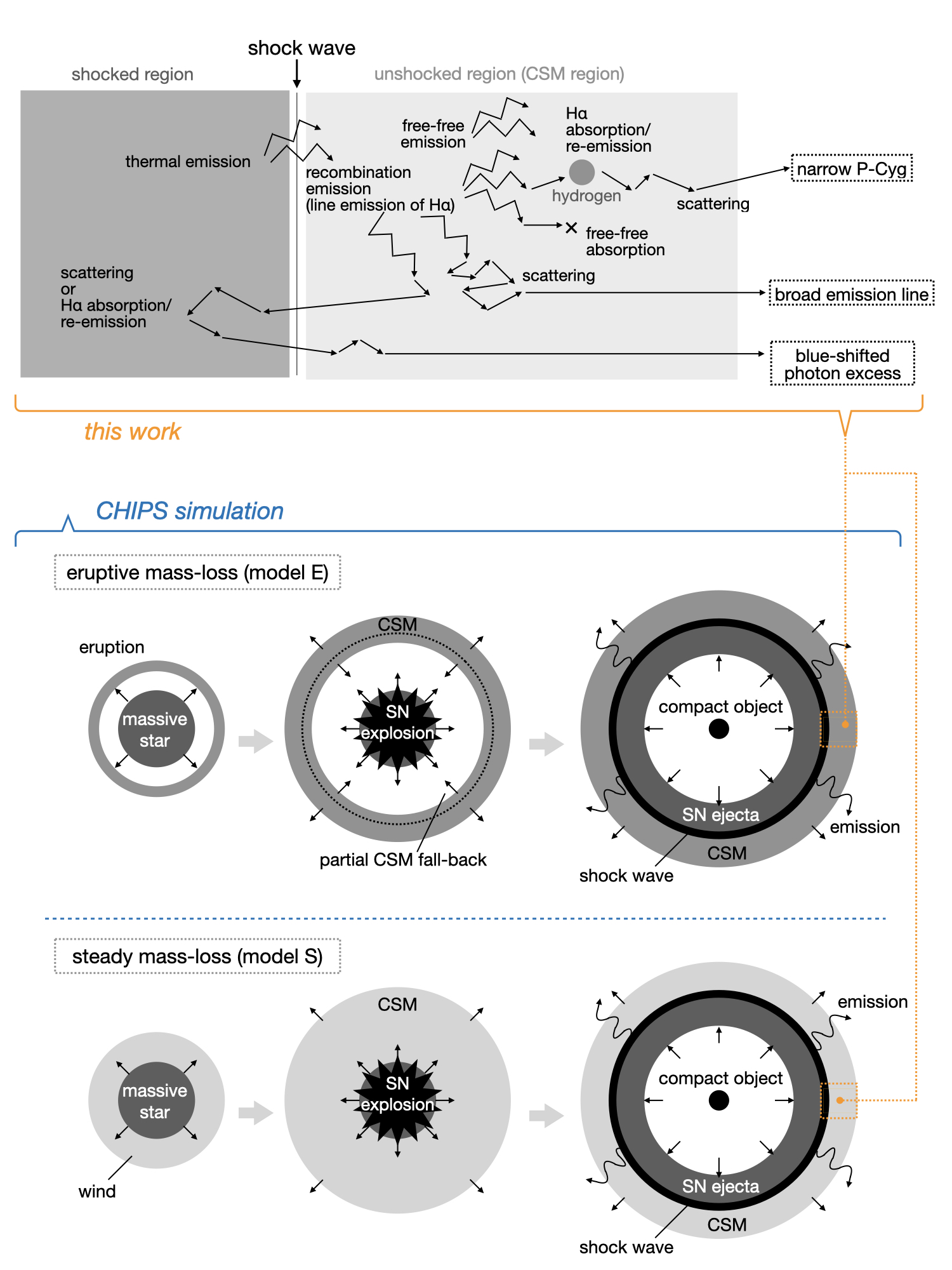}
    \caption{Schematic picture of the emission from interacting SNe that we model in this work. 
    The upper side shows the line emission computed by Monte Carlo radiative transfer simulations (this work), and lower side shows the radiation hydrodynamical part computed by the CHIPS code.}
    \label{fig:abstract}
\end{figure}

Our model for spectral calculations solves the radiative processes around the forward shock front that separates the shocked gas and the surrounding unshocked dense CSM, as depicted in the upper side of Figure \ref{fig:abstract}. 
The structure of each region as a result of ejecta-CSM interaction is first obtained using the CHIPS code \citep{takei2022,Takei_et_al_2023}, which has been demonstrated to be able to simulate both the formation of the dense CSM and the subsequent interaction between the SN ejecta and the CSM.

Using CHIPS we perform radiation hydrodynamics simulations of the interaction between the ejecta and the dense CSM, with the latter considering two scenarios of enhanced mass-loss, eruptive or steady mass loss. 
Using the hydrodynamical outputs from CHIPS as background, we then perform post-process Monte-Carlo simulations to obtain the spectrum at each epoch, focusing on the features of the H$\alpha$ emission. The model setup for each case is described in detail below.

\subsection{Simulating Type IIn Supernovae by CHIPS}
\label{sec:chipsmodels}
We first obtain a representative SN progenitor model with the stellar evolution code \verb|MESA| version 12778 \citep{Paxton_2011,Paxton_2013,Paxton_2015,Paxton_2018,Paxton_2019,Jermyn23}. We use the \verb|example_make_pre_ccsn| test suite, and evolve until core-collapse a non-rotating star with zero-age main-sequence mass of $M_\mathrm{ZAMS}=20M_\odot$. The chemical compositions of hydrogen, helium and metals at ZAMS are set to be $X=0.718$, $Y=0.268$ and $Z=Z_\odot(=0.014)$ respectively.
The resultant progenitor is a red supergiant (RSG) that has a photospheric radius of $R_*=843R_\odot$, luminosity of $L_*=1.75\times10^{5}L_\odot$, and effective temperature of $T_\mathrm{eff}=4062\,{\rm K}$. We have adopted the ``Dutch" wind mass-loss rates of \cite{Nieuwenhuijzen90}, which has reduced the mass of the progenitor to $M_*\approx18.3M_\odot$ upon core-collapse. 
As the structure of the hydrogen envelope hardly changes in the last decades of its life, we adopt the progenitor at core-collapse as the progenitor before enhanced mass loss. 

\subsubsection{Models of the Dense Circumstellar Material}
We consider two scenarios of the enhanced mass loss from this progenitor that forms the dense CSM, via eruptive (model E) and steady mass loss (model S). 
The lower side of Figure \ref{fig:abstract} shows a schematic picture of the evolution of a SN in a dense CSM for both of the two models. 
A massive star generates the dense CSM through enhanced mass loss prior to an SN explosion. Once the massive star explodes as an SN, the SN ejecta collides with the CSM, generating shock-powered emission along with the  narrow lines from reprocessing in the CSM. 
The remainder of this sub-subsection describes the settings shared by model E and model S, and the settings that differ for each model, respectively.

\begin{itemize}

\item Shared parameters for both models

\noindent
Some model parameters characterizing the CSM and SN ejecta share the same values in both models for ease of comparison. The main parameters that characterize the dynamics of the CSM interaction are the mass and radial extent of the CSM. As described in detail below, we adopt the same extent of $r_{\rm CSM}\approx 3\times 10^{15}$ cm and total mass of $M_{\rm CSM}\approx 1.8M_\odot$. The subsequent SN is assumed to leave a neutron star remnant of $M_{\rm rem}=1.4\ M_{\odot}$. 
The SN ejecta is assumed to have a kinetic energy of $10^{51}\ \rm{erg}$, 
with ejecta mass of $M_*-M_{\rm CSM}-M_{\rm rem}\approx15.1 M_{\odot}$. Following common prescriptions \citep{Matzner99}, the ejecta is set to be homologous with a double power-law density profile (see equation 6 of \citealt{takei2022}), with outer and inner power-law indices of $n = 11.43$ and $\delta = 1$ respectively.

\item Eruptive mass loss (model E)

\noindent    
We generate the CSM with the mass eruption module of the CHIPS code \citep[for details see][]{kuriyama2020}, by injecting energy into the base of the star's outer envelope and following its response. We inject an energy of $6.0\times10^{47}\,{\rm erg}$, which is 80\% of the envelope's binding energy.
The SN is assumed to happen 10 years after the energy injection, which results in a dense CSM of $M_{\rm CSM}\approx 1.8\ M_{\odot}$ with a dense part extending out to $r_{\rm{CSM}}\approx 3\times 10^{15}$ cm. The values chosen generally reproduce the light curve morphology (rise time and peak luminosity) typically seen in SNe IIn \citep{takei2022}. The velocity of the resulting CSM monotonically increases with radius, with the velocity in the outermost part of the dense CSM of $\approx 100$ km s$^{-1}$ which is comparable to the progenitor's escape velocity.
Before the eruptive mass loss, the star should have been losing mass by blowing a steady wind. To include this, we smoothly connect a stellar wind component out to $r_{\rm out}=3\times 10^{16}$ cm the same as \cite{takei2022}, with a mass-loss rate of $\dot{M} \approx 1.1 \times 10^{-5}\ M_{\odot}\rm{/yr}$ and velocity of $\approx 24$ km/s. 
These values are obtained from the formulae in \cite{takei2022} and are rather arbitrary, but the density in this region is so small that it would not affect the results of the subsequent radiative transfer simulations.

\item Steady mass loss (model S)

\noindent
The progenitor is supposed to suddenly start blowing a much denser wind than before, and continue blowing at a constant mass-loss rate till the progenitor explodes. We adopt a double power-law density profile for the CSM, similar to the shape of model E but with the inner region of a wind profile of $\rho\propto r^{-2}$,
\begin{eqnarray}
    \rho_\mathrm{CSM}(r) = \rho(r=r_{\rm CSM})\left[\frac{(r/r_\mathrm{CSM})^{n_\mathrm{in}/3.5}+(r/r_\mathrm{CSM})^{n_\mathrm{out}/3.5}}{2}\right]^{-3.5},
\end{eqnarray}
where the inner and outer power-law indices are set to $n_\mathrm{in}=2$ and $n_\mathrm{out}=12$, respectively. Similar to model E, we smoothly connect a tenuous stellar wind of mass-loss rate $\dot{M}\approx 1.1 \times 10^{-5}\ M_{\odot}\rm{/yr}$. 
The normalization for $\rho$ is chosen so that the total CSM mass is the same as model E. 
The largest difference from model E is the initial velocity profile; the CSM velocity is assumed to be 24 km s$^{-1}$ independent of radius. This is slower than the outermost part of the dense CSM in model E by a factor of 4, and thus corresponds to the enhanced mass loss starting $\approx 40$ years before core-collapse\footnote{The wind can have different density and velocity profiles close to the stellar radius $R_*$ if wind acceleration is taken into account \citep[e.g.,][]{Moriya18}. While this difference can be important for the emission $\lesssim 10$ days after explosion, our interest is in the spectral features after this epoch when the shock has already swept up the CSM at $r\gtrsim R_*$.}. 

\end{itemize}

\subsubsection{Modelling the Circumstellar Interaction}
With these CSM profiles as inputs, we model collisions with the SN ejecta using the SN module of the CHIPS code \citep[for details see][]{Takei20,takei2022}. The code is split into two steps, the shock dynamics and radiative transfer in the unshocked CSM. The former solves the propagation of the forward and reverse shocks formed by the collision, as well as the detailed structure of the shocked region taking into account radiative transfer in the flux-limited diffusion approximation \citep{Levermore81}. The radiative flux at the forward shock, emitted outwards in the unshocked CSM, is simultaneously obtained. 

With the time-evolving forward shock luminosity as input, the second step solves radiation transfer in the unshocked CSM that is assumed to be a two-temperature fluid, where the gas and radiation temperatures separately evolve with time. The number of computational cells in the unshocked CSM is set to 1000 at the beginning of the interaction of the SN ejecta with the CSM, 
and decreases with time as the CSM is engulfed by the shock. The width of the narrow lines, which is characterized by the velocity of the unshocked CSM, can be affected by acceleration due to radiation produced from shock interaction \citep{Chevalier11,Tsuna23}.
We thus add radiative acceleration $a$ of the unshocked CSM at each radii $r$ as $a(r)=\kappa(r) L(r)/4\pi r^2c$, 
where $\kappa(r)$ is the total (scattering + absorption) opacity adopted in the CHIPS code, $L(r)$ is the incoming luminosity, and $c$ is the speed of light. The opacities in the calculation are referenced from opacity tables, generated for (mass-averaged) hydrogen, helium, and metal abundances in the CSM of $X=0.6402$, $Y=0.3457$, and $Z=0.014$ respectively. 

\begin{figure*}
   \centering
    \begin{tabular}{cc}
     \begin{minipage}[t]{0.5\hsize}
    \centering    \includegraphics[width=\linewidth]{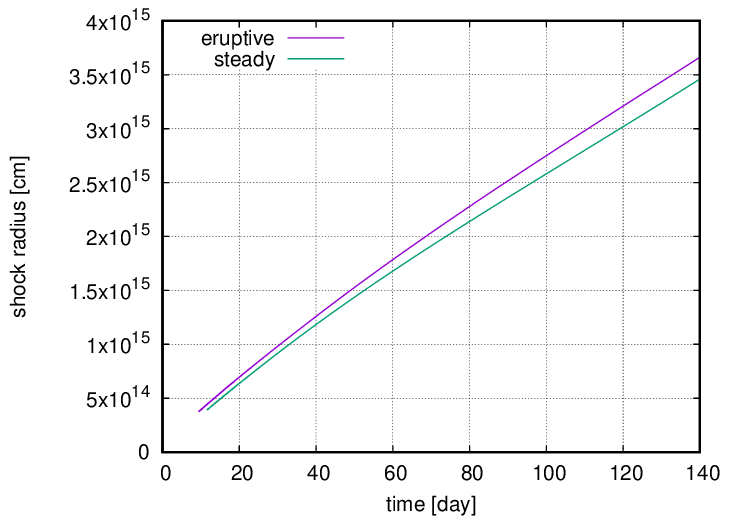}
    \end{minipage}
    
     \begin{minipage}[t]{0.5\hsize}
    \centering
    \includegraphics[width=\linewidth]{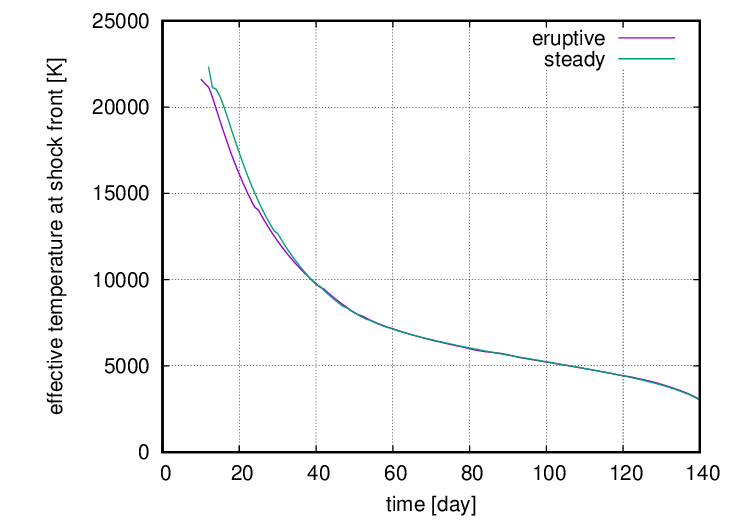}
    \end{minipage} \\

    \begin{minipage}[t]{0.5\hsize}
    \centering
    \includegraphics[width=\linewidth]{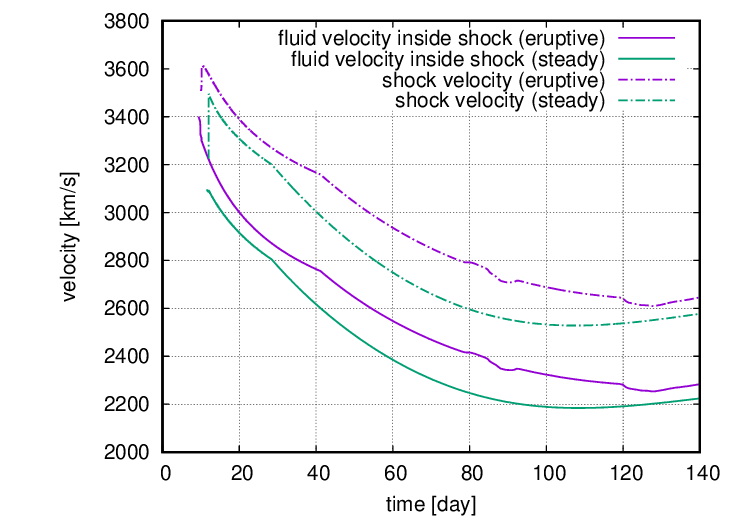}
    \end{minipage}

     \begin{minipage}[t]{0.5\hsize}
    \centering
    \includegraphics[width=\linewidth]{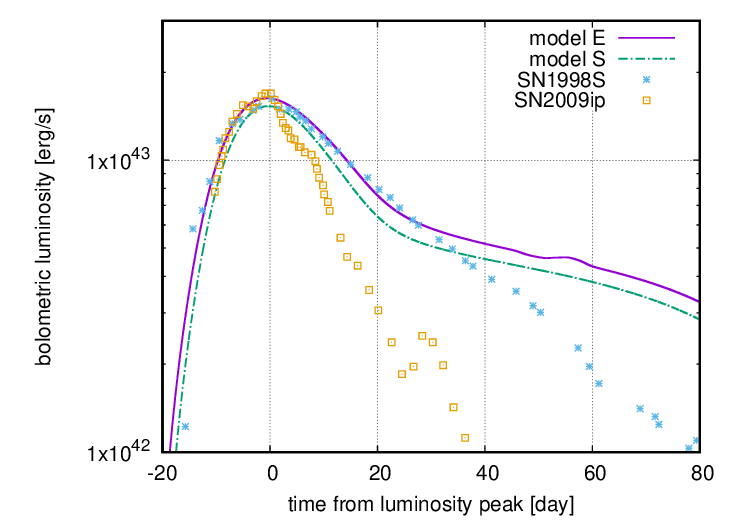}
    \end{minipage}   
    \end{tabular}
    \caption{The time-dependent forward shock radius (top-left), effective temperature at the shock front (top-right), the fluid velocity inside the shock wave accompanied with the forward shock velocity (bottom-left), and bolometric light curves (bottom-right) for models E and S. We also show for reference the light curves of two well-known Type IIn SNe, 1998S and 2009ip. The x-axis of the light curves are shifted so that the peak of luminosity is at $t=0$.}
    \label{fig:shock_info}
\end{figure*}

\begin{figure}
    \centering
    \includegraphics[keepaspectratio, scale=0.3]{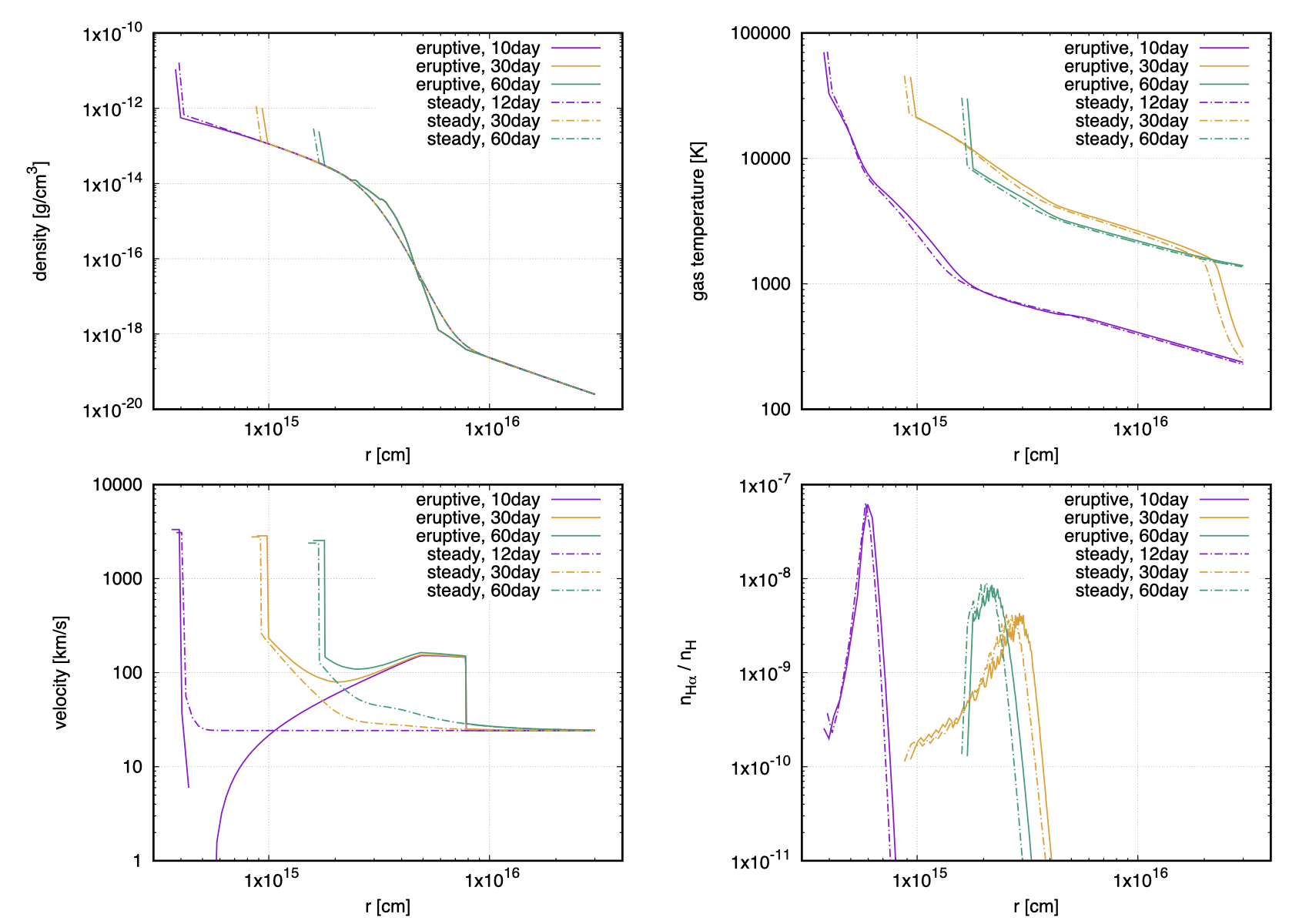}
    \caption{The distributions of gas density (top-left), gas temperature (top-right), velocity (bottom-left), and the ratio of number densities of excited hydrogen that causes H$\alpha$ absorption among total hydrogen (bottom-right) 
    of the shocked region and the CSM at 10/12 (model E/S, respectively), 30, and 60 days after the the SN explosion. 
    For model E at day 10, the unshocked CSM at $r\approx 5\times 10^{14}$ cm is falling back with a negative velocity, hence does not appear in the bottom-left panel. 
    }
    \label{fig:ini_setup}
\end{figure}
Figure \ref{fig:shock_info} shows the temporal evolution of the forward shock generated by the interaction: the shock radius $r_{\rm fs}$, the effective temperature at $r=r_{\rm{fs}}$, and the shock velocity together with the fluid velocity at the immediate downstream. 
The shock in model E has a higher velocity than in model S due to the slightly lower density in the innermost region of the CSM. 
In the bottom-left panel of Figure \ref{fig:shock_info}, shock velocity increases at the earliest time for both models and then begins to monotonically decrease, which is due to the initial setting of the CHIPS simulations. 

The bottom-right panel of Figure \ref{fig:shock_info} shows the bolometric light curves of models E and S generated by CHIPS, measured at $r_{\rm out}=3\times 10^{16}$ cm. The CSM interaction generally dominates the light curve when the shock is running through the dense CSM. As the dense CSM in these two models has similar mass and extent, the resulting light curves are similar through day 80. The light curves from our models have rise time ($\approx 15$ days) and peak luminosity ($\approx 10^{43}$ erg s$^{-1}$) that are representative in the parameter space of SN IIn \citep{Ofek14_rise_time_peak,Nyholm20}, and are similar to the well-known SNe like 1998S \citep{Fassia2000MNRAS.318.1093F} and 2009ip \citep{Margutti14} overlaid in the plot. 

Figure \ref{fig:ini_setup} shows distributions in the shocked region and the CSM of gas density, velocity, gas temperature, and the number density ratio $n_{\rm{H\alpha}}/n_{\rm{H}}$ of excited hydrogen that causes H$\alpha$ absorption among the total hydrogen, at days 10/12 (start of the spectrum simulation for model E/S), 30, and 60 after the SN explosion.
The ratio $n_{\rm{H\alpha}}/n_{\rm{H}}$ is calculated assuming local thermodynamical equilibrium, with temperature equal to the gas temperature obtained from the CHIPS simulation. 
The number density $n_{\rm{H\alpha}}$ in the fluid comoving frame of excited hydrogen atoms with $n = 2$ that can cause H$\alpha$ is given as 
\begin{equation}
    \label{eq:n_ha}
    n_{\rm{H\alpha}} = n_{\rm{HI}} \frac{g_{\rm{2}} {\rm exp}(-h \nu_2 /k_{\rm B} T)}{\Sigma_{n=1}^\infty g_n {\rm exp}(-h \nu_n/ k_{\rm B} T)},
\end{equation}
where $n_{\rm{HI}}$ is the number density of neutral hydrogen atoms in the fluid comoving frame, $g_n$ is the statistical weight of the level $n$, $\nu_n$ is the frequency corresponding to the energy difference between level $n$ and the ground level, and $T$ is the gas temperature. 
We actually take the summation in the denominator of the right hand side only from $n=1$ to $n=3$. As explained in detail in the end of Section \ref{sec:method}, the degree of ionization is referenced from a pre-computed table. This creates small fluctuations of $n_{\rm{H\alpha}}/n_{\rm{H}}$ as can be seen in the figure, but this is found to have a negligible effect on the subsequent radiative transfer calculations. 
 
The shock breaks out at 10 days from explosion in model E, while it breaks out on 12 days in model S because of its slightly larger optical depth of the CSM.
As Figure \ref{fig:ini_setup} shows, while the unshocked CSM in both of the models have similar density structures, they exhibit different velocity distributions. Such different velocity profiles are preserved even after the CSM is radiatively accelerated by photons from the interaction. As demonstrated in Section \ref{sec:result}, this difference results in distinct features of the H$\alpha$ spectra in the two models.

\subsection{Post-process Monte-Carlo Radiative Transfer Simulations}
\label{sec:method}

The CHIPS code computes light curves from interaction-powered SNe, but it cannot compute detailed spectra. 
To obtain them, we perform post-process radiative transfer simulations assuming steady-state background flow-field, using the snapshots of the density, velocity and temperature profiles in the shocked region and the unshocked CSM from Section \ref{sec:chipsmodels}. 
The steady-state assumption is valid during the post-breakout interaction phase, as the hydrodynamical properties do not significantly change within the diffusion timescale of photons in the relevant regions, except for the early phase soon after the interaction of SN with CSM occurs.

The computational domain and radiation processes included in the simulation are shown in the upper part of Figure \ref{fig:abstract}. 
We illustrate what kind of radiation process is important in which part of the computational domain, and how each process affects the line shape of the resulting spectrum (details are described in Section \ref{sec:result}). 
Photons from the shock front are assumed to be emitted as blackbody radiation, since the photons in the optical (and UV) wavelength range are expected to be immediately thermalized in the shocked CSM with a large mass-loss rate of $\dot{M}\gtrsim 0.1M_\odot\ {\rm yr^{-1}}\ (v_{\rm CSM}/100\ {\rm km\ s^{-1}})$ \citep{Chevalier_Irwin_12,Svirski12,Tsuna21_IIn,Margalit22}. We further incorporate hydrogen recombination emission and free-free emission in the unshocked CSM. 
Photons experience the collisional processes such as electron scattering, H$\alpha$ absorption and re-emission, and free-free absorption. 
We use a Monte Carlo method for radiation transfer simulations \citep{lucy1999, lucy2005, maeda2014, ishii2017}. 
The fundamental part of photon transport and electron scattering process are following \cite{ishii2017}, in which transport in a relativistic flow-field is considered by taking into account Compton scattering, while thermal motion of electrons is ignored. In this study, the Doppler effect by the thermal motion of electrons following a Maxwell distribution with gas temperature is taken into account for scattering because this effect is important in determining the line shape. In addition the recombination emission, free-free emission/absorption, and line absorption/re-emission processes are installed in this study. 
A ``packet" of photons with the same frequency is treated as a sample particle. We use a total of $10^8$ packets for calculating the spectrum at each epoch, which is verified to be enough for convergence. 
 
In this study we focus only on the line shape of H$\alpha$, and thus the initial photon frequency distribution is limited to the vicinity of the H$\alpha$ transition frequency. 
Packets for the continuum photons are initially assigned to a frequency in the range of $0.9 \nu_{\rm{H\alpha}} \leqq \nu_{\rm{0}} \leqq 1.2 \nu_{\rm{H\alpha}}$   
in the fluid comoving frame, where $\nu_{\rm{H\alpha}}=c/(656.28~{\rm nm})$ 
is the H$\alpha$ photon frequency, while packets for H$\alpha$ photons via recombination emission process are assigned to $\nu_{\rm{0}} = \nu_{\rm{H\alpha}}$. 
Even if the energy distribution in the fluid comoving frame is limited to around H$\alpha$, it is broadened in the observer frame depending on the photon direction and the fluid velocity. 
Since a large number of photon packets is required to resolve the characteristic narrow lines in interaction-powered SNe as well as the broadened (often called ``intermediate-width") lines, we need to focus on such narrow energy range to save the computational cost. 

The computational region consists of the shocked and unshocked CSM. As the profiles of the physical quantities in the shocked region are found to be nearly constant in radius, we treat this region as one zone for simplicity. At a given time, we determine quantities of this zone by averaging the radial profile of the shock dynamics calculations in CHIPS. 
First, the average density $\bar{\rho}_{\rm{s}}$ and internal energy $\bar{u}_{\rm{s}}$ are calculated by a volume weighted average in the shocked region. The average temperature $\bar{T}_{\rm{s}}$ is then calculated from 
\begin{eqnarray}
\bar{u}_{\rm{s}} = \frac{3}{2} \frac{\bar{\rho}_{\rm{s}}}{\mu m_{\rm{u}}} k_{\rm{B}} \bar{T}_{\rm{s}} + a_{\rm{r}} \bar{T}_{\rm{s}}^4, 
\end{eqnarray}
where $\mu(\bar{\rho}_{\rm s}, \bar{T}_{\rm s})$ is the mean molecular weight obtained by solving Saha equations with hydrogen and helium, $m_{\rm{u}}$ is the atomic mass unit, $k_{\rm B}$ is the Boltzmann constant, and $a_{\rm{r}}$ is the radiation constant, respectively. 
The average velocity $\bar{v_{\rm{s}}}$ is obtained by dividing the total momentum of the shocked region by its total mass.

The width of this zone $\Delta r_{\rm{s}}$ is determined so that $\Delta r_{\rm{s}} = \Delta \tau_{\rm{s}}/\bar{n}_{\rm{e}} \sigma_{\rm{T}}$, where $\sigma_{\rm{T}}$ is the Thomson scattering cross section and $\Delta \tau_{\rm{s}}$ is the Thomson scattering optical depth in the shocked region derived by integration over the region using the density and temperature profiles obtained from CHIPS.
The one-zone electron number density $\bar{n}_e(\bar{\rho}_s, \bar{T}_s)$ is calculated by solving Saha equations for mixture of hydrogen and helium gas assuming thermal equilibrium. 
We obtain the ionization degrees of hydrogen and helium by referencing from a pre-computed table for given values of $n_{\rm{HI}}$, the number density for neutral helium atoms $n_{\rm{HeI}}$ in the fluid comoving frame, and gas temperature $T$. 
The values of $n_{\rm{HI}}$, $n_{\rm{HeI}}$, and $T$ in the table cover a broad range of $10-10^{15}\ \rm{cm^{-3}}$, $0.1-5\times10^{13}\ \rm{cm^{-3}}$, and $10^{3}-10^{5}$ K, respectively.  
Here, the mass fractions for hydrogen and helium are set to be the same as the CHIPS calculations, and the contribution to $n_{\rm{e}}$ from heavier elements is neglected for simplicity.

\subsubsection{Emission processes}

We assume that the radiation emitted from the forward shock front is a blackbody, and thus emit a packet of photons of frequency $\nu$ with a luminosity 
\begin{equation}
   L_{\nu} =4\pi r_{\rm fs}^2B_\nu(T_{\rm eff,s})= \frac{8 \pi r_{\rm fs}^2h\nu^3}{c^2\left[\exp(h\nu/k_{\rm B}T_{\rm eff,s})-1\right]},
\end{equation}
where $B_\nu$ and $h$ are respectively the Planck function and the Planck constant, and $T_{\rm eff,s}$ is the effective temperature at the shock front. 
Thermal photons from the shocked region are initially located just $10^{-6}$ \% outside of $r_{\rm{fs}}$ and are emitted outwards. 
For each Monte-Carlo simulation, both the duration of emitting photons and sampling photons exiting the edge of the computational region at $r_{\rm out} = 3 \times 10^{16}$ cm are set to $2 \times 10^6$ s. We set this by the sum of the diffusion time for our CSM models ($\approx 10^6$ s) and the light-crossing time in the computational region $r_{\rm out}/c=10^6$ s.\footnote{We focus on the spectra at $\gtrsim 20$ days from explosion, so the hysteresis effect due to photon diffusion is expected to be insignificant.}

Soon after CSM interaction begins, hydrogen atoms in the CSM are ionized by radiation from the shocked region, and recombines with free electrons to emit the H$\alpha$ line as well as other hydrogen lines. The transition from $n>1$ to $n=1$ can be neglected, because the emission from such transition is immediately absorbed (case B recombination; \citealt{osterbrock2006}) for a very dense environment like the CSM in SNe IIn.
The case B emission coefficient $j_{\rm{H\alpha}}(n_e,T)\ \rm{[erg\  cm^{-3}]}$ of the recombination emission in H$\alpha$ is obtained from Table 4.4 of \cite{osterbrock2006}. 
Since the original table is coarse, we interpolate the values of $j_{\rm H\alpha}$ as a function of $n_e$ and $T$. The ranges of $n_e$ ($10^2$--$10^6$ cm$^{-3}$) and $T$ (5000--20000 K) in the table do not fully cover the ranges required in this study, so we linearly extrapolate this table to the values of $n_e$ and $T$ requested\footnote{The coefficient very weakly depends on $n_e$, with changes of only 0.5-5$\%$ for a 4 dex change in $n_e$, so the value would not depend much on the choice of extrapolation. 
Although the coefficient is more sensitive to $T$, the CSM temperature exceeds 20000 K only in the innermost region and only shortly after the explosion; the temperature in the entire CSM drops to less than $20000$ K from about 30 days after explosion.
}. 
The H$\alpha$ photons are isotropically emitted, with their initial positions determined from a probability distribution proportional to the product of the emission coefficient and the volume at each computational cell. 

The frequency $\nu$ of the photon packet with free-free emission process in the CSM is determined by the spectral emission coefficient due to bremsstrahlung \citep[e.g.][]{zeldovich1967}. 
We consider hydrogen and helium for the ions and neglect the contribution from heavier elements. 
Similar to the recombination emission, photons are emitted isotropically with their initial positions following the product of the frequency-integrated emission coefficient and the volume at each computational cell.

\subsubsection{Transport process}
\label{subsec:line_emi_abs}

To investigate the spectral shape of the H$\alpha$ line from the CSM with some velocity gradients, we use the Sobolev approximation as introduced in \cite{sobolev1960} and used in many previous works \citep[e.g.,][] {tanaka2013,maeda2014,kerzendorf2014, dessart2015}. 
While the velocity gradient in the unshocked CSM is much smaller than the SN ejecta, the part of the CSM that contributes to H$\alpha$ absorption is also radiatively accelerated (see Figure \ref{fig:ini_setup}), which generally creates a large enough velocity gradient to justify this assumption.

The optical depth $\tau_{\rm{H \alpha}}$ for H$\alpha$ absorption in the observer frame is obtained as (for details see Appendix \ref{app:tau_ha}), 
\begin{equation} \label{eq:tauha}
    \tau_{\rm{H \alpha}} = \frac{\pi e^2}{m_e c} f n_{\rm{H\alpha}} \left(\frac{d\nu}{ds}\right)^{-1},
\end{equation}
where $f$ is the oscillator strength of the transition, $d\nu/ds$ denotes the frequency change along the photon path in the  observer frame. If we denote the angle between the path and the radial direction $\theta$ and the velocity distribution $v(r)$, the term of $d \nu / ds$ is written as 
\begin{equation}\label{eq:dnuds}
    \frac{d\nu}{ds}=-\nu_{\rm H\alpha}\frac{d}{dr}\left(\frac{v(r)\cos\theta}{c}\right)\frac{dr}{ds}.
\end{equation}
Since the velocity $v(r)$ in the relevant region is smaller than a few thousand km/s, the terms of the order of $\mathcal{O}(v/c)^2$  or higher are negligible. 
In actual calculations, $(d\nu/ds)^{-1}$ is obtained from equation (\ref{eq:dsdnu}) along a finite path $\delta s$ the photon travels. 

To simulate the trajectory of each photon packet, we consider the mean free path of each scattering/absorption process. The mean free path of photons for H$\alpha$ transition, electron scattering, and free-free absorption are given as
\begin{equation}
    l_{\rm{H \alpha}} = \frac{1}{\frac{\pi e^2}{m_e c} f n_{\rm{H\alpha}} \frac{1}{\delta \nu}},\ l_{\rm{es}} = \frac{1}{\sigma_{\rm{T}} n_{\rm{e}}},\ l_{\rm{ff}} = \frac{1}{\rho \kappa_{\nu}},
\end{equation} 
respectively, where  $\kappa_{\nu}$ is the free-free absorption coefficient, which can be obtained from the spectral emission coefficient, $j_{\rm{ff,}\nu}$, and Kirchhoff's law $j_{\rm{ff,}\nu} = \kappa_{\nu} B_{\nu}$. 
For $l_{\rm H\alpha}$, the line profile function is given by the $\delta$-function in the Sobolev approximation where line broadening by the thermal motion of atoms is not taken into account. This can result in $\delta \nu = 0$, which is unphysical. To avoid that, $\delta \nu$ is assumed to have a minimum value of $3\times 10^{-5} \nu_{\rm{H\alpha}}\ \rm{s^{-1}}$, which corresponds to the velocity of thermal motion for hydrogen atoms at $T \sim 10^4$ K. 
Since the floor value of $\delta\nu$ might change the depth of the absorption lines seen in the spectrum only if the velocity gradient is negative everywhere, we have to note what assumptions are used to determine the floor value. We will discuss this issue in section \ref{sec:spectrum_ha_profile}.

For a single step of a photon packet, we determine which reaction occurs using a random number $R$ in the range of $0\leq R\leq 1$ so that the probability of each reaction is inversely proportional to its mean free path. The actual distance $l$ the packet travels is determined by another random number so that $R=\exp(-l/l_{\rm sc}-l/l_{\rm ff}-l/l_{\rm{H \alpha}})$. 
If a packet experiences H$\alpha$ absorption, it is re-emitted at the same position with an energy of $\nu_{\rm{H\alpha}}$ in a random direction in the fluid comoving frame. 
When a packet is determined to be absorbed by a free-free transition of an ion, the packet is erased at the position.
Our treatment of line radiative transfer is validated in Appendix \ref{app:jeffery}, with a test problem of modeling the P-Cyg line emerging from a homologously expanding medium \citep{jeffery1990}.

\section{Results} \label{sec:result}
In this Section we present spectra calculated for models E and S introduced in section \ref{sec:chipsmodels} at several epochs up to $t=80$ days, where the time is defined as $t=0$ at explosion.  

\subsection{Spectra of H$\alpha$ line profiles}
\label{sec:spectrum_ha_profile}

\begin{figure}
    \centering
    \includegraphics[keepaspectratio, scale=0.3]{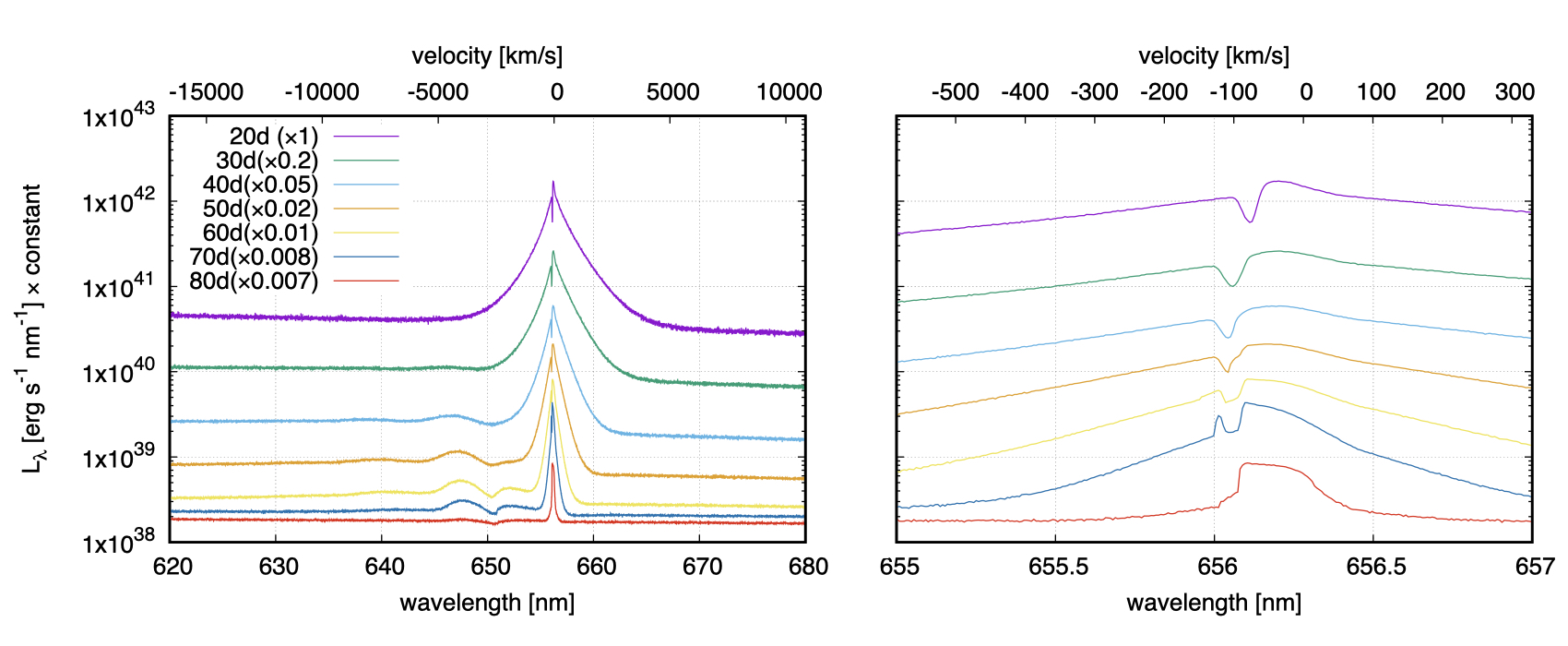}
    \caption{Time-dependent spectra in the eruptive mass-loss model (model E). For better visibility, the spectra after 20 days are multiplied by constants as indicated on the legend. The right panel is an enlarged view of the left panel to show the narrow lines around the H$\alpha$.}
    \label{fig:spec_time_eruptive}
\end{figure}

Figure \ref{fig:spec_time_eruptive} shows snapshots of spectra around the H$\alpha$ wavelength for model E. 
The spectra generally show a combination of a narrow P-Cyg line and a broad emission line, with their shapes determined by the CSM structure immediate outside the shock wave. 
The CSM in the vicinity of the shock front is significantly accelerated due to the effect of radiation in the early phase ($t\lesssim30$ d in Fig. \ref{fig:ini_setup}), while the CSM velocity in the vicinity of the shock front does not significantly change as the radiation flux decreases in the later phase ($t\gtrsim60$ d). 
The wavelength of the absorption minimum in the P-Cyg profile likely traces the minimum velocity of the CSM affected by the radiation from the shock, as seen from the right panel of Figure \ref{fig:spec_time_eruptive}. 

The H$\alpha$ emission line is broadened by a large number of electron scattering in the CSM, as previously claimed in \cite{chugai2001} and \cite{huang2018}. The wavelength shift $\Delta\lambda$ when a line photon incident from a direction $\vec n$ is scattered in a direction $\vec n_1$ is given as
 \begin{equation}\label{eqn:shift}
 \Delta\lambda\sim\frac{\lambda_{{\rm H}\alpha} \vec p\cdot(\vec n-\vec n_1)}{m_{\rm e}c}=\frac{\lambda_{{\rm H}\alpha}p(\cos\theta-\cos\theta_1)}{m_{\rm e}c},
 \end{equation}
 where $\lambda_{{\rm H}\alpha}$ is the wavelength of H$\alpha$ emission line in the rest frame of the hydrogen atom, $\vec p$ denotes the momentum vector of an electron \citep{rybicki1979}, and $p=|\vec p|$. Here $\theta$ and $\theta_1$ are the angles between the momentum vector of the electron and the vectors $\vec n$ and $\vec n_1$, respectively. 
 Since photons travel a distance $s = \tau_{\rm es}^2/(n_{\rm e}\sigma_{\rm T})$ on average in a CSM with an optical depth $\tau_{\rm es}(>1)$ for electron scattering before diffusing out of the CSM, the line width can be estimated as
 \begin{equation}
 \label{eq:dlambda}
\sqrt{\langle\Delta\lambda^2}\rangle/\lambda_{{\rm H}\alpha}\sim\tau_{\rm es}\sqrt{\frac{2k_{\rm B}T}{m_{\rm e}c^2}}\sim2\times10^{-3}\tau_{\rm es}\sqrt{\frac{k_{\rm B}T}{1\ {\rm eV}}}.
\end{equation}
The right hand side can be expressed as $\sqrt{y/2}$ in terms of the Compton $y$ parameter \citep{rybicki1979}. Comparison of equation (\ref{eq:dlambda}) and the line widths in the left panel of Figure \ref{fig:spec_time_eruptive} indicates that H$\alpha$ photons are emitted in a region with an optical depth $\tau_{\rm es}$ of at most a few, even in the early phases such as day 30 when the total Thomson optical depth of the unshocked CSM is $\approx 20$. 
This is because photons emitted by H$\alpha$ transitions in deeper regions are scattered by electrons and mostly change back close to the H$\alpha$ frequency to be absorbed by the same H$\alpha$ transitions, and hence cannot contribute to the observed H$\alpha$ emission line.
As the shock expands, the temperature and ionization degree in the CSM also decrease. 
Then the number of electron scatterings and the Compton $y$ parameter decrease, resulting in a narrower emission line. 

\begin{figure}
    \centering
    \includegraphics[keepaspectratio, scale=0.3]{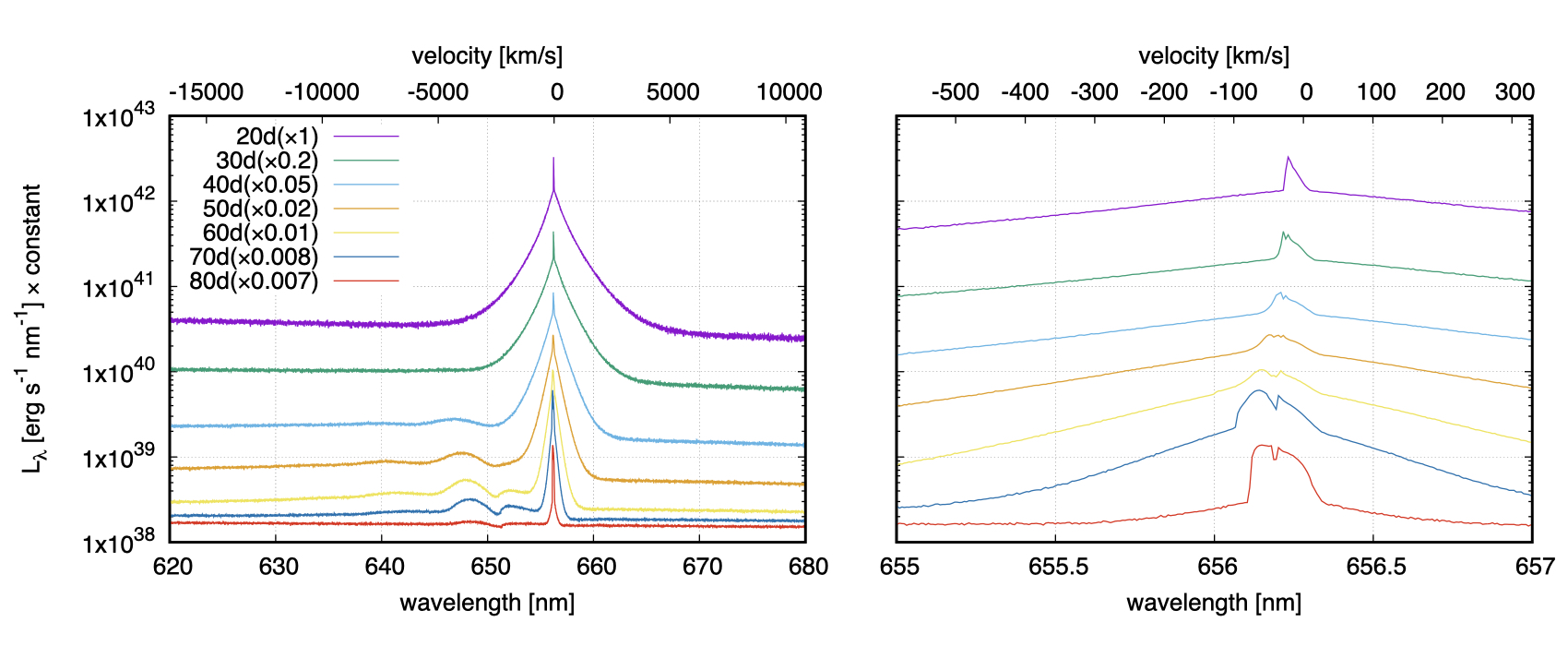}
    \caption{Same as Figure \ref{fig:spec_time_eruptive}, but for the steady mass-loss model (model S).}
    \label{fig:spec_time_steady}
\end{figure}

Figure \ref{fig:spec_time_steady} shows the time evolution of the spectrum for model S. 
At the initial phase, the line profile is composed of a combination of broad and narrow emission lines as in model E. The time evolution of the broad component is almost the same as model E, as this only depends on the time-varying $\tau_{\rm es}$ of the CSM that is similar between the two models. However, the shape of the narrow lines appear differently, with no clear absorption lines as seen in model E.

As seen in the right panel of Figure \ref{fig:spec_time_steady}, a small absorption feature appears at the velocity of $\sim-30$ km s$^{-1}$ in the broad emission component after day 30. 
This absorption line is expected to appear from a medium with a negative velocity gradient, and the similar issue was already noted in \cite{dessart2015}. In model S the velocity gradient is negative everywhere, in contrast to model E in which the velocity increases with radius at the outer part of the dense CSM (the bottom left panel of Fig. \ref{fig:ini_setup}), keeping the memory of the eruption. If the velocity gradient is negative, there is a direction for which the frequency shift ($\delta\nu$ in Eq. (\ref{eq:dnuds})) vanishes. 
In terms of the angle $\theta$ from the radial direction, this condition can be expressed as 
\begin{equation}\label{eq:pecd}
    \frac{d\ln v}{d\ln r}=-\tan^2\theta
\end{equation}
because $d\nu\propto d(v\cos\theta)=\left(dv/ds\cos\theta-v\sin\theta d\theta/ds\right)ds=0$, $d\theta/ds=-\sin\theta/r$, and $dr/ds=\cos\theta$ along the path $ds$ making an angle $\theta$ with the radial direction. 
This means that a photon emitted  in this direction by the H$\alpha$ transition is immediately absorbed and re-emitted in a different direction by the H$\alpha$ transitions and results in a strong absorption feature in the spectrum.
 In reality, the thermal motion of hydrogen atoms prevents $\delta\nu$ from vanishing and instead increases $\delta\nu$ of the photon emitted in the direction satisfying equation (\ref{eq:pecd}) up to a value of the order of $\nu_{\rm{H\alpha}}\sqrt{2k_{\rm B}T/m_{\rm p}c^2}$ at minimum. Thus we set the lower limit of $3 \times 10^{-5} \nu_{\rm{H\alpha}}\ s^{\rm{-1}}$ to $\delta\nu$, which corresponds to $\nu_{\rm{H\alpha}}\sqrt{2k_{\rm B}T/m_{\rm p}c^2}$ at $T \sim 10^4$ K, to mimic this situation as already mentioned in Section \ref{subsec:line_emi_abs}. The wavelength of the absorption feature traces the wind velocity as seen in Figure \ref{fig:spec_time_steady} and does not significantly change with time. This is in contrast with model E, in which the wavelength of the absorption feature moves to shorter wavelength with time.

\subsection{Comparison of spectra for different CSM models }

\begin{figure*}
   \centering
    \begin{tabular}{cc}
     \begin{minipage}[t]{0.5\hsize}
    \centering
    \includegraphics[width=\linewidth]{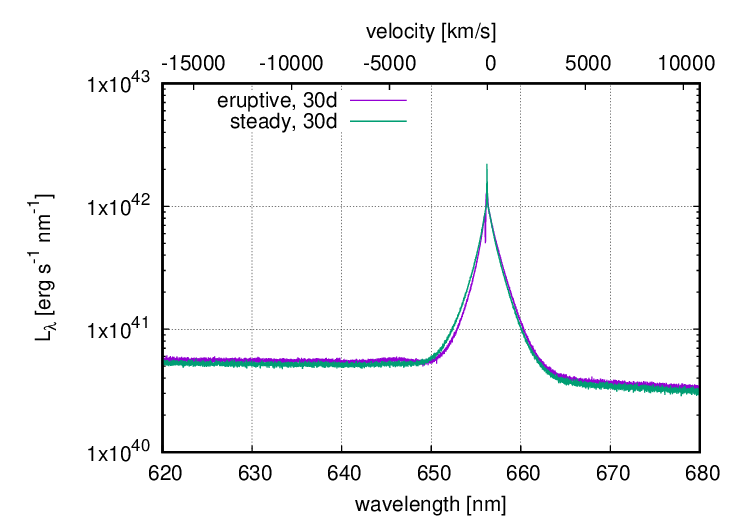}
    \end{minipage}
     \begin{minipage}[t]{0.5\hsize}
   \centering
    \includegraphics[width=\linewidth]{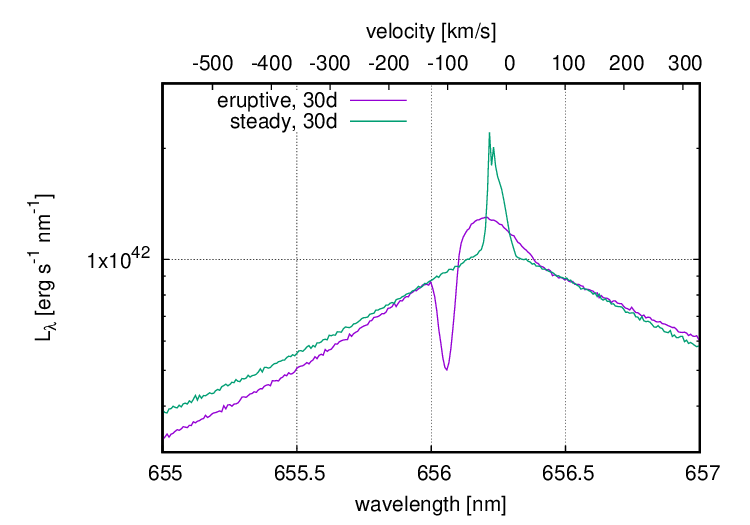}
    \end{minipage}
    \end{tabular}
    \caption{Comparison of the H$\alpha$ line shapes between models E and S at day 30.}
    \label{fig:compare_eruptive_steady}
\end{figure*}
\begin{figure*}
   \centering
    \begin{tabular}{cc}
     \begin{minipage}[t]{0.5\hsize}
    \centering
    \includegraphics[width=\linewidth]{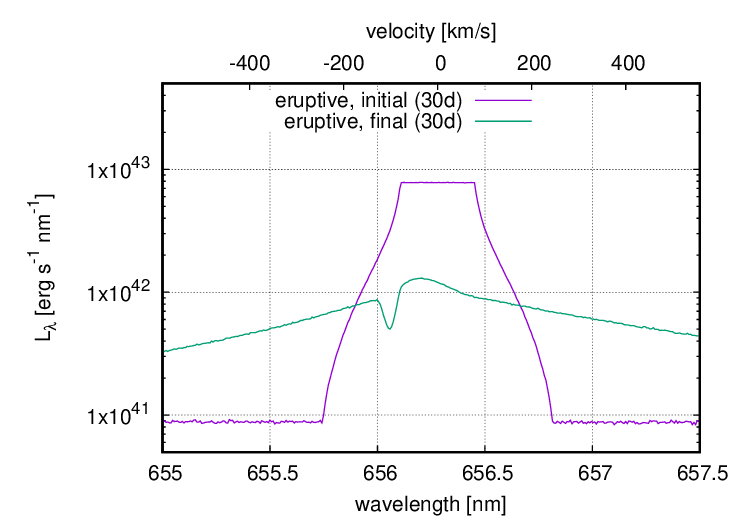}
    \end{minipage}
     \begin{minipage}[t]{0.5\hsize}
   \centering
    \includegraphics[width=\linewidth]{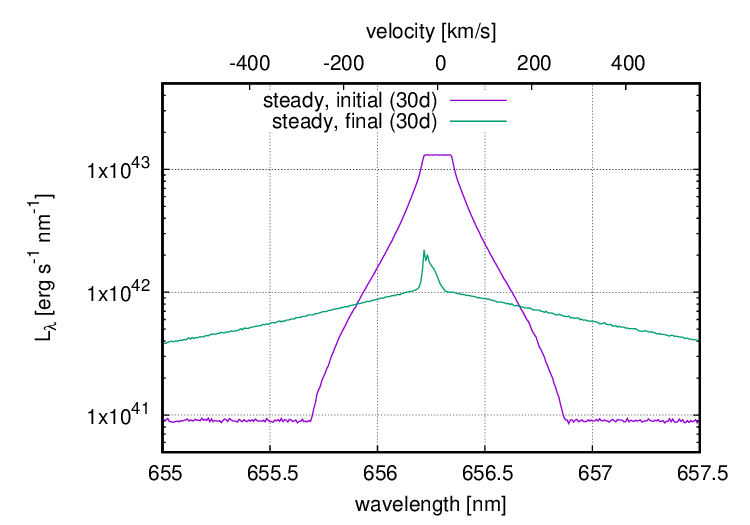}
    \end{minipage}
    \end{tabular}
    \caption{Comparison of the initial and final spectra at the day 30. Left:model E. Right: model S.}
    \label{fig:emit_escape}
\end{figure*}

Figure \ref{fig:compare_eruptive_steady} shows a comparison of the the spectra calculated for the two models at day 30. 
As mentioned in the previous section, while the broad lines are similar in the two models (left panel) the narrow lines appear very different (right panel). Model E shows a P-Cyg like profile with a narrow absorption line at $v\approx -100$ km s$^{-1}$, while for model S no such deep absorption line appears. To understand what causes this difference, we overplot in Figure \ref{fig:emit_escape} the ``initial" spectra of photon packets emitted at the start of calculation. Here, all emitted photons are sampled for the ``initial" spectra, whereas the observed spectra (denoted as ``final" in Figure \ref{fig:emit_escape}) are composed only of photons escaping from the outer boundary. 

In both models, the ``initial" spectra have a symmetrically broadened structure with respect to the wavelength of H$\alpha$ in the rest frame. The H$\alpha$ emission line is produced by recombination of hydrogen, and the broadening of the line up to $\sim300$ km s$^{-1}$ indicates that H$\alpha$ photons are emitted in the vicinity of the shock front where the CSM is accelerated by radiation from the shocked region. The width of the plateau in the ``initial" spectra is determined by the minimum velocity in the ionized region, which is $\sim80$ km s$^{-1}$ in model E and $\sim30$ km s$^{-1}$ in model S. 
Most of these photons experience electron scattering, free-free absorption, and absorption and re-emission by the H$\alpha$ transitions. The luminosity of the emergent emission line component decreases by a factor of $\sim$10 due to these processes, and the line is further broadened due to electron scattering as discussed above. The emergent spectrum has a blue-shifted peak in both models. This is because a significant fraction of photons entering the inner boundary of the computational region (i.e. shocked region), which are red-shifted in the observer frame, do not contribute to the emergent spectrum. The lack of these redder photons in the observed spectrum creates the blue-shifted peak. 

As we describe below, the difference in the detailed structure of the emission line for the two models is due to the difference in the velocity gradient of the unshocked CSM (Figure \ref{fig:ini_setup}). In model S, a narrow emission line appears at around $-$30 km s$^{-1}$, which is the bluest edge of the plateau in the ``initial" spectra. This narrow emission line is composed of photons that are emitted in nearly radial directions, and escape from the CSM with no further reactions. 
In model S, the velocity gradient is negative at all locations, due to a combination of a flat initial velocity profile and the radiative acceleration being stronger at smaller radii. With such a velocity distribution, when a photon is initially emitted with $\nu_{\rm{H\alpha}}$ in the fluid comoving frame and travels in the radial direction, it would never have the frequency $\nu_{\rm H\alpha}$ again unless it experiences electron scattering. 
Since $L_\lambda$ of this emission line is reduced by a factor of $\sim6$ from the ``initial" value, we infer that the emitting region of this line has an electron scattering optical depth of $\tau_{\rm e}\sim \ln(6)\approx 1.8$. 
This value is roughly consistent with $\tau_{\rm{es}}$ estimated by equation (\ref{eq:dlambda}). 

\begin{figure*}
   \centering
    \begin{tabular}{cc}
     \begin{minipage}[t]{0.5\hsize}
    \centering
    \includegraphics[width=\linewidth]{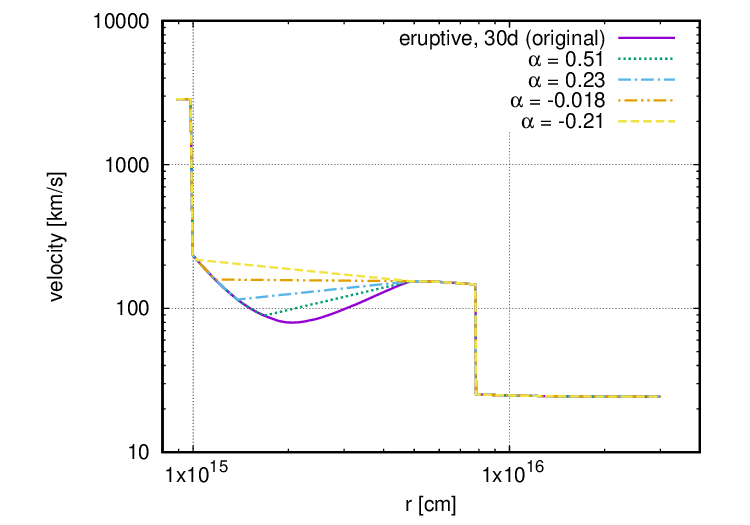}
    \end{minipage}
     \begin{minipage}[t]{0.5\hsize}
   \centering
    \includegraphics[width=\linewidth]{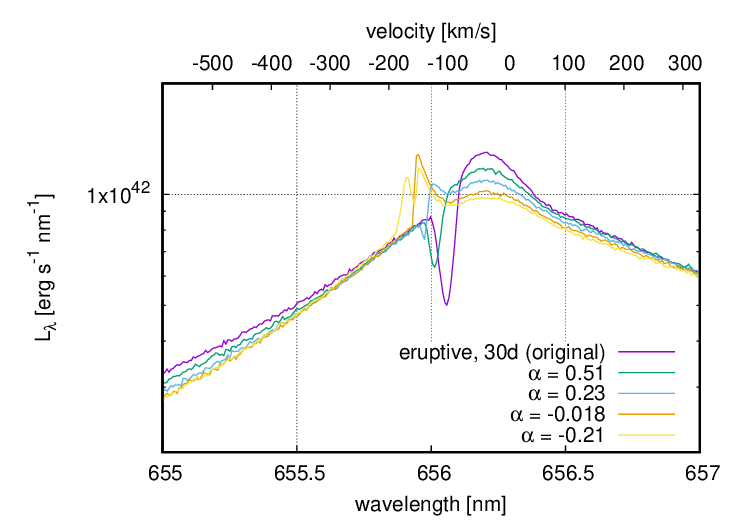}
    \end{minipage}
    \end{tabular}
    \caption{Comparison of the line shapes at day 30 for various modified velocity distributions in the unshocked CSM, assumed to be power-law with the indices varied as in the left panel. The solid line shows the original velocity profile for the eruptive model.}
    \label{fig:velo_abs_line_check}
\end{figure*}

To further demonstrate this relationship between the velocity structure and the narrow line feature, we perform Monte-Carlo calculations of several models in which only the velocity distribution is artificially modified from model E while the density and temperature distributions are fixed (Fig. \ref{fig:velo_abs_line_check}). 
For these modified models we assume that the CSM velocity from each location in the outside of the shock wave to $\sim5\times 10^{15}$ cm has a power law distribution $v \propto r^{\alpha}$, with $\alpha>0\ (\alpha<0)$ corresponding to a positive (negative) velocity gradient as seen in the left panel of Figure \ref{fig:velo_abs_line_check}. 
The right panel of Figure \ref{fig:velo_abs_line_check} shows the corresponding spectra for the assumed velocity distributions in the left panel. 
When $\alpha$ is positive, a larger $\alpha$ makes the absorption line deeper.  
On the other hand, when $\alpha$ takes a negative value, the absorption line disappears and an emission line appears. These results confirm that the velocity gradient explains the difference in the line profiles in Figure \ref{fig:emit_escape}.
A similar trend of distinct absorption features in H$\alpha$ line produced by CSM with different velocity distributions was pointed out by \citet{chugai2004}, which modeled a particular SN IIn 1994W. They claimed that a homologously expanding CSM produces prominent absorption lines, while what we show here is that CSM with a positive velocity gradient produces prominent absorption lines and the part of the CSM contributing to the H$\alpha$ line formation does not expand homolgously in our model.

\subsection{Blue-shifted excess component}
\label{subsec:blue_excess}

\begin{figure}
    \centering
    \includegraphics{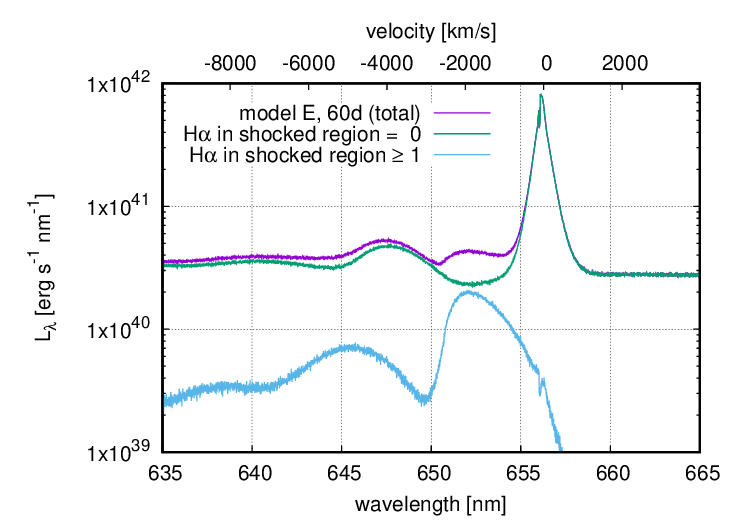}
    \caption{Spectrum and the components decomposed by the number of H$\alpha$ absorption and the subsequent re-emission processes in the shocked region for model E at day 60.}
    \label{fig:ha-shock}
\end{figure}

\begin{figure}
    \centering    \includegraphics{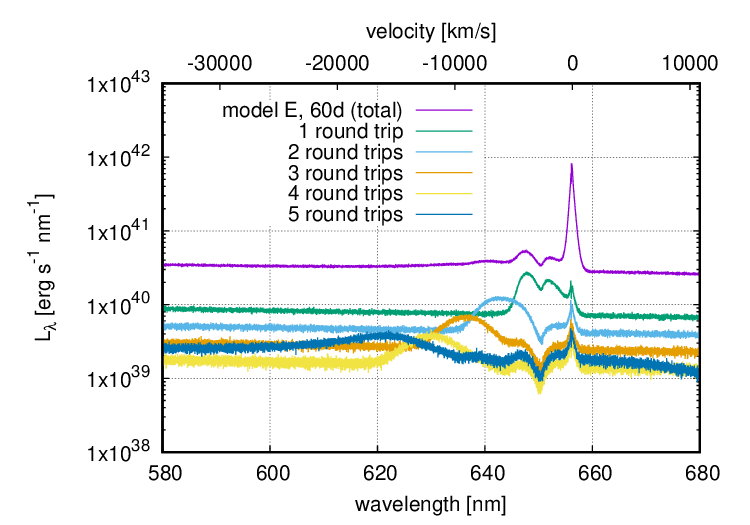}
    \caption{Spectrum and the components decomposed by the number of photon round-trips across the shock wave, for model E at day 60. Here a round-trip means that photons initially emitted in the unshocked CSM cross the shock wave, are scattered by electrons or absorbed and re-emitted by H$\alpha$ transition in the shocked region, and return to the unshocked CSM. The ``5 round trips" are photons that experience 5 or more round trips. 
    }
    \label{fig:spec_shock}
\end{figure}

Returning to the time-dependent spectra in Figures \ref{fig:spec_time_eruptive} and \ref{fig:spec_time_steady}, an excess from the continuum emission is seen on the blue side at around 650 nm. 
The blue-shifted excess component emerges as a bump at around day 40, while from around day 60 a smaller and redder component appears between the main emission line and the first blue component. 
From inspection of the photon trajectory, we find that the bluer (redder) excess is formed by H$\alpha$ photons from recombination in the unshocked CSM, which enters the shocked region and is scattered by electrons (absorbed and re-emitted by the H$\alpha$ transition) there, and exits the shocked region towards the observer without further reprocessing. Figure \ref{fig:ha-shock} shows the spectrum and its components decomposed by the number of H$\alpha$ absorption/re-emission in the shocked region for model E at day 60. The second redder component appears if the number of H$\alpha$ absorption/re-emission in the shocked region is larger than 1, while the component does not appear when the H$\alpha$ process does not occur in the shocked region. Thus, we see that the second redder component is caused by photons that experience the H$\alpha$ process in the shocked region.

If a photon in the shock downstream encounters an electron moving with the downstream bulk flow and changes its direction closer towards the radial direction by scattering, the photon acquires energy from the electron in the observer frame as bulk Compton scattering process and results in the wavelength shift according to Equation (\ref{eqn:shift}). Such photons produce the first bluer excess, with a maximal wavelength shift of
\begin{eqnarray}
\Delta\lambda_{\rm max, scat} \approx \frac{\lambda_{\rm H\alpha}\cdot 2m_ev_{\rm down}}{m_e c} = \frac{2v_{\rm down}\lambda_{\rm H\alpha}}{c},
\label{eqn:lambda_max_scat}
\end{eqnarray}
where $v_{\rm down}$ is the downstream bulk velocity.
If a photon in the downstream is instead absorbed and re-emitted by H$\alpha$ transitions, the photon re-emitted in a nearly radial direction would also have a higher energy in the observer frame due to the bulk motion of the shock downstream. The resultant wavelength shift is at most, 
\begin{equation}
\label{eq:lambda_max_ha}
\Delta \lambda_{\rm max, H\alpha} \approx v_{\rm down}\lambda_{\rm H\alpha}/c,
\end{equation}
half of the wavelength shift due to electron scattering. This is why two excess components appear at the blue shifted wavelengths.

To quantitatively see the effect of shock crossing on the photon spectra, we show in Figure \ref{fig:spec_shock} the spectra at day 60, grouping the photons with the number of round-trips across the shock wave. The position of the blue-shifted excess seen in the total spectrum (e.g. left panel of Fig. \ref{fig:spec_time_eruptive}) is mainly reproduced by photons with one round-trip. 
Photons experiencing two or more round trips make an even bluer and broader component, with the amount of blue-shift proportional to the number of the round trips. However the wavelength of the redder component hardly changes with the number of round-trips. This is because photons with multiple round-trips via electron scattering in the shocked region would gain energy each time they are scattered, whereas photons with round-trips via H$\alpha$ absorption/re-emission would always have a frequency of $\nu_{\rm{H\alpha}}$ in the comoving frame upon re-emission. 
Since the number of photons significantly decrease for higher numbers of round trips, the component with one round trip mostly determines the total spectrum. Nevertheless, one may be able to see multiple blue-shifted excess components due to multiple round-trips, for a spectrum of a bright SN IIn with a good resolution and photon count.

We note that the second redder excess is sensitive to the temperature in the shocked region as it involves H$\alpha$ transitions. 
In this particular model, the temperature of the shocked region between days 60 and 70 is $T\approx 3\times10^4$ K, just around the temperature where hydrogen could cause H$\alpha$ transitions. If the shock wave were stronger and the temperature of the shocked region were higher, the second excess component would disappear. 
Thus while the second component may be sensitive to the assumed shock downstream temperature, the main bluer excess seen in our simulations should be robust.

The blue-shifted excess is evident only at late times, as the electron scattering optical depth in the unshocked CSM has to be sufficiently low for the excess to be notable.
At early times, the large $\tau_{\rm{es}}$ spreads photons in this blue-shifted excess component to a wider wavelength range, which smooths out this component and makes it buried either in the broad line or the continuum. 

\begin{table}[t]
    \centering
    \caption{The fluid velocity inside the shock wave and the minimum wavelengths of the photon with electron scattering or H$\alpha$ absorption/re-emission inside the shock wave.}
    \label{tab:scat_photon_vs}
    \begin{tabular}{c|c|c|c}
    \hline
          Time from Explosion (days)
          & velocity [km/s] & minimum $\lambda$ for scattering [nm] & minimum $\lambda$ for H$\alpha$ transition [nm] \\
    \hline
    \hline
    40   & 2763.8 & 644.17 & 650.22 \\
    50   & 2646.2 & 644.69 & 650.48 \\
    60   & 2547.0 & 645.12 & 650.70 \\
    70   & 2469.0 & 645.47 & 650.87 \\
    \hline
    \end{tabular}
\end{table}

\begin{figure}
    \centering
    \includegraphics[scale=0.3]{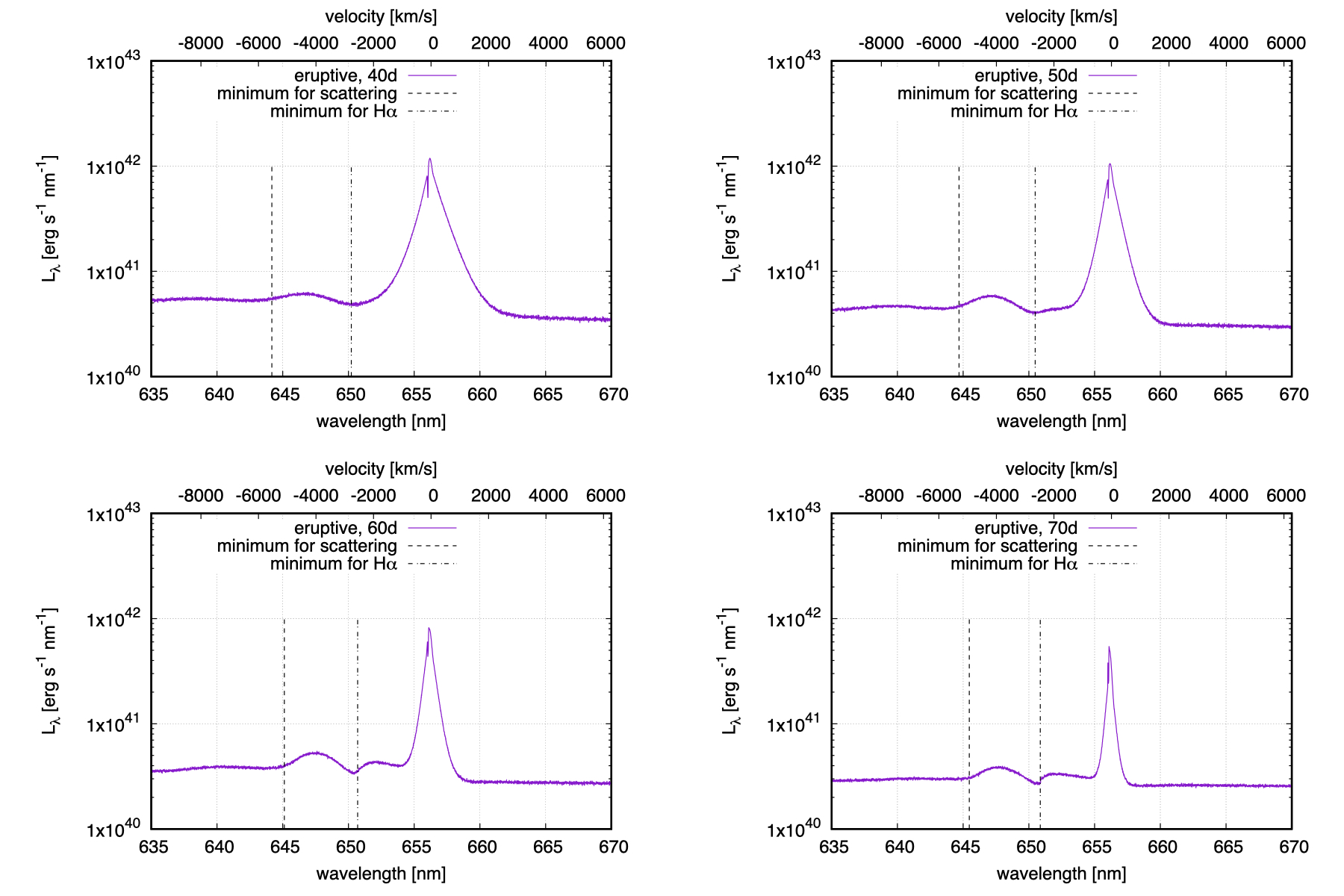}
    \caption{Spectra including the blue-shifted photon excess. The minimum wavelengths of the photon with electron scattering or H$\alpha$ transition inside the shock wave are shown with the vertical lines.  
    Top-left: the spectrum at 40 day. Top-right: the spectrum at 50 day. Bottom-left: the spectrum at 60 day. Bottom-right: the spectrum at 70 day.} 
    \label{fig:blue_bump_velo}
\end{figure}

Table \ref{tab:scat_photon_vs} shows the velocity inside the shock wave and the corresponding minimum wavelengths for the photon with electron scattering or H$\alpha$ absorption/re-emission inside the shock wave obtained with equations (\ref{eqn:lambda_max_scat}) and (\ref{eq:lambda_max_ha}). 
Figure \ref{fig:blue_bump_velo} shows each spectrum from day 40 to day 70 with vertical lines indicating the minimum wavelengths. Since these lines successfully trace the bluer edges of the two blue-shifted components, one can estimate the velocity of the shocked matter and its evolution, if either of the two components is clearly seen in the spectra. 
\footnote{Note that the above calculations of $\Delta \lambda$ do not take into account the shift due to the electron thermal velocity, which may slightly shift the vertical lines in Figure \ref{fig:blue_bump_velo} to the blue side.}

\section{discussion and conclusion}
\label{sec:conclusion}

In this study, we modelled spectra emitted from interaction of SN ejecta with an extended dense CSM, formed by either eruptive or steady mass-loss prior to the SN, using our newly developed Monte Carlo radiative transfer code. We focused on the H$\alpha$ line, and investigated in detail the physical processes that cause the broad component and narrow P-Cyg profile, and the blue-shifted excess component which have been observed in some SNe IIn.  

As summarized in the top panel of Figure \ref{fig:abstract}, our model predicts three features in the spectra: the broad component, narrow component, and the blue-shifted excess component at late phases. The broad component is formed by H$\alpha$ photons emitted by the recombination  experiencing a large number of electron scattering in the CSM, consistent with the claims of \cite{chugai2001} and \cite{huang2018}.
The narrow P-Cyg profile is formed by H$\alpha$ photons experiencing a small number of electron scattering and the subsequent H$\alpha$ absorption and re-emission in the CSM.
The blue-shifted excess is caused by H$\alpha$ photons interacting with the matter in the shocked region, which then travel almost radially to the unshocked CSM. 
The excess can have two components, with interaction in the shocked region being electron scattering or H$\alpha$ absorption/re-emission corresponding to a bluer or redder component, respectively. 
This excess does not appear until the optical depth for electron scattering in the unshocked CSM becomes small enough, which happens at around day 40 for our particular model. 

Comparing the spectral shapes of the H$\alpha$ line between the CSM models formed in the two scenarios, the narrow P-Cyg line is only observed for the CSM modelled by eruptive mass loss.
We find that this is due to the difference in their velocity distributions of the CSM in the pre-SN phase. 
For CSM formed by steady mass-loss with constant velocity profile, the velocity after SN monotonically decreases with radius due to radiative acceleration. A photon initially emitted from the H$\alpha$ transition cannot undergo H$\alpha$ absorption unless the traveling direction satisfies Equation (\ref{eq:pecd}) or it changes its direction and wavelength via electron scattering. 
Otherwise the photon could never again have the wavelength of H$\alpha$ in the comoving frame of any other hydrogen in its path.
Therefore, the absorption line is absent in the steady mass-loss case, while it appears in the eruptive mass-loss case in which the velocity distribution includes a weak shock-like structure with positive velocity gradient. 
We confirm the relation of the absorption line feature and velocity gradient through test calculations using artificially modified velocity distributions and find that the absorption line appears when there is a positive velocity gradient, and that a steeper positive velocity gradient deepens the absorption line.  

For both CSM models a blue-shifted excess appears in the spectra between days 40 and 70. This is caused by electron scattering and H$\alpha$ absorption/re-emission in the shock downstream, corresponding to the first bluer excess and second redder one, respectively. From this finding, we analytically estimated the wavelength changes of photons by electron scattering or H$\alpha$ transition using the flow velocity at the immediate downstream. 
The analytical estimates of the wavelengths at the bluer edges of the two blue-shifted excesses agree with the numerical results at all times. 

From our relationship between the H$\alpha$ line shape and the CSM structure, we claim that the CSM structure and its origin can be diagnosed by high-resolution spectroscopic observations of H$\alpha$ lines. 
Since the presence of a narrow absorption line is important as evidence for the eruptive mass-loss, a resolution of $\lambda/\Delta\lambda>c/v_{\rm CSM}\sim10^4(v_{\rm CSM}/30\ {\rm km\ s^{-1}})^{-1}$ is required to distinguish the CSM-forming mechanisms. Here $v_{\rm CSM}$ denotes the typical velocity of the unshocked CSM in the interaction region, and is dependent on the progenitor. While we considered here an RSG progenitor, the requirement on resolution would likely be less stringent for CSM formed from more compact stars, such as LBVs with expected $v_{\rm CSM}\gtrsim 100\ {\rm km\ s^{-1}}$.

\subsection{Applications to Observed SNe}

\begin{figure*}
    \centering
    \begin{tabular}{c}
    \begin{minipage}[t]{0.5\hsize}
    \centering
    \includegraphics[width=\linewidth]{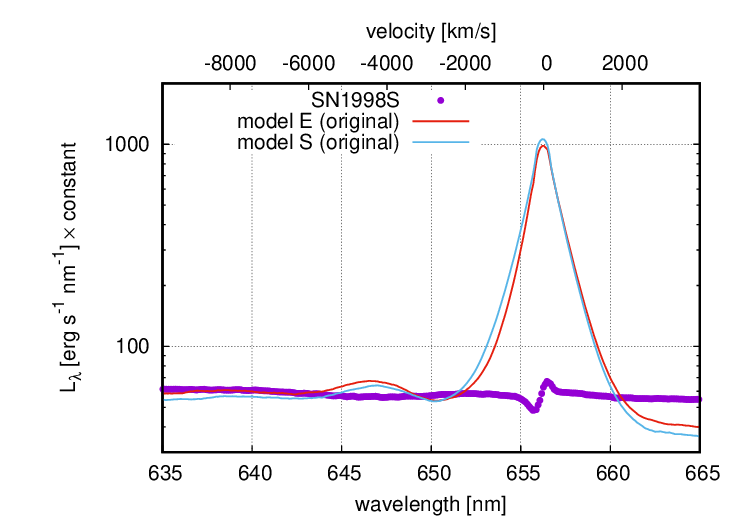}
    \end{minipage}
    \end{tabular}
   \centering
    \begin{tabular}{cc}
     \begin{minipage}[t]{0.5\hsize}
    \centering
    \includegraphics[width=\linewidth]{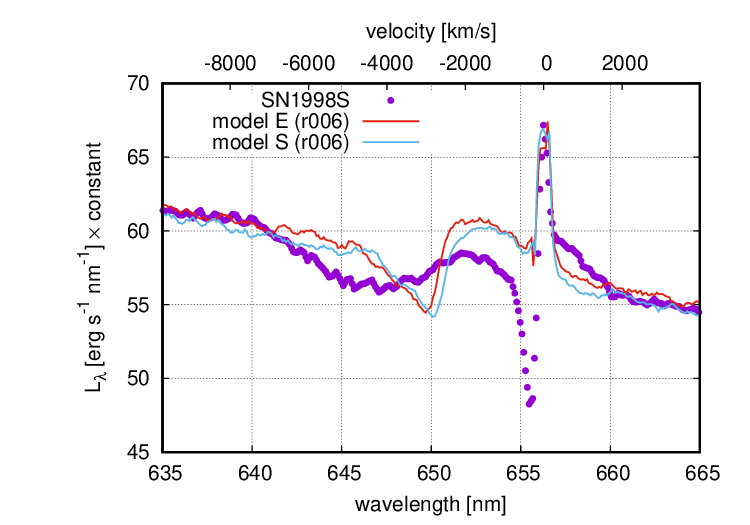}
    \end{minipage}
     \begin{minipage}[t]{0.5\hsize}
   \centering
    \includegraphics[width=\linewidth]{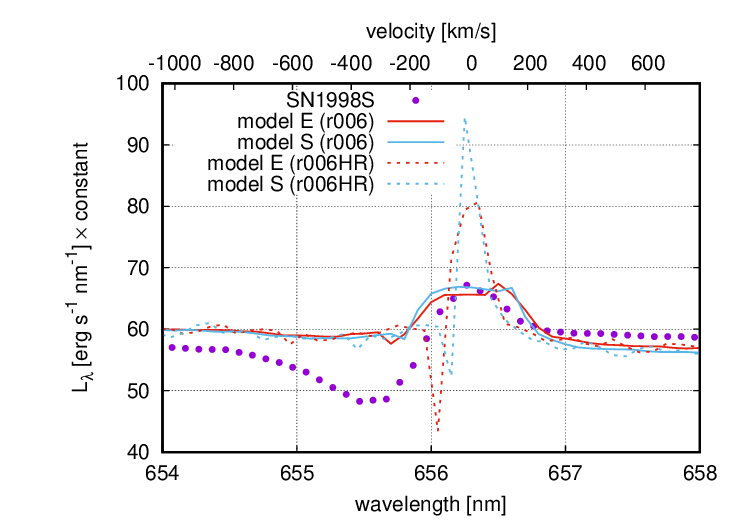}
    \end{minipage}
    \end{tabular}
    \caption{Comparison of the observed spectrum of SN 1998S and the simulation results 40 days after the SN explosion for both models E and S. Spectra from our simulations are constructed by binning photon packets to an equal bin-width of $\Delta \lambda = 0.8$ nm and shifting each of the bins by 0.1 nm to be consistent with the observed spectrum. Top: Comparison with the simulation results of the original models of E and S. Bottom-left: Comparison with the models of decreased CSM density by a factor of 0.06.  Bottom-right: Comparison of observation with simulated spectra with different bin widths. `r006' means the same as the bottom-left panel and `r006HR' means high-resolution spectra with $\Delta \lambda =0.1$ nm.}
    \label{fig:compare_obs}
\end{figure*}

Our work is aimed to be an in-depth study of the various spectral features we expect from interaction-powered SNe, and we thus did not aim to fit a particular observed spectrum. Nevertheless, our findings can lead to some qualitative understanding of the CSM environment for some of the observed SNe. As a preliminary demonstration, we show a comparison between the observed spectrum of SN 1998S and our simulation results. We also apply our findings to two SNe IIn, SN 1997ab \citep{salamanca1998} and KISS-15s \citep{kokubo2019}.

Since our models E and S give rising times and peak luminosities consistent with those observed in SN 1998S as shown in Figure \ref{fig:shock_info},  we also compare the spectra. We obtain the spectral data from WISeREP \citep{Yaron12} and focus on the spectrum observed 9 days after the luminosity peak by Keck1/LRIS. Spectra from our simulations are constructed by binning photon packets to an equal bin-width of $\Delta \lambda = 0.8$ nm and each bin is shifted by 0.1 nm for rebinning to be consistent with the observation.  
Figure \ref{fig:compare_obs} shows a comparison of the observed spectrum of SN 1998S and the simulation results 40 days after the SN explosion for both models E and S. 
This period is chosen because it roughly corresponds to 9 days after the luminosity peaks in models E and S. As shown in the top panel of this figure, the original simulation results in a too strong H$\alpha$ emission line. 
On the other hand, the bottom-left panel shows the simulation spectra from models with the same CSM structures but with a reduced density distribution by a factor of 0.06, in which the density only in the unshocked region is reduced and that in the shocked region is unchanged,
 roughly reproduces the strength of the observed line. 
To create a less dense CSM at 40 days after the SN explosion without changing the shape of the light curve, there can be the following suggestion: shorten the time duration between eruption and SN explosion while maintaining the density of the inner dense part of the CSM that powers the light curve around peak, 
which creates a model where the shock radius is further outward and CSM density is lower at day 40. Since some different parameter sets result in similar light curves in CHIPS calculations, a detailed parameter survey is required to create a variety of flow-field profiles without changing the light curve, which will be clarified in the future work. 

As shown in the bottom-right panel of Figure \ref{fig:compare_obs}, the difference in line shapes between models E and S is not pronounced in the low-resolution spectrum while it is more significant in the high-resolution spectrum. Thus, high-resolution spectra are necessary to distinguish CSM structures formed from different mass-loss histories. 
The difference in the width of the narrow absorption line between the observation and simulations may be partly due to the limited wavelength resolution of the observation, which is indicated that the full width at half-maximum of the narrow line is less than 300 km/s \citep{Leonard00}. Furthermore, it seems difficult to simultaneously explain the depth of the absorption line and the height of the emission line seen in the observed spectrum with a spherically symmetric CSM model, because deeper absorption lines require higher CSM density, which leads to too strong recombination emission to be compatible with the observation. To match both absorption and emission lines between the observation and simulations, a non-spherically symmetric CSM model 
 with high density only around the line of sight of the observer is required. 

As the bottom-left panel shows, the dip around 640 -- 650 nm in the observed spectrum is qualitatively reproduced by our simulations. As described in Section \ref{subsec:blue_excess}, the dip is located between the two photon-excess components created by electron scattering and H$\alpha$ absorption/re-emission inside the shock wave, and the characteristic wavelengths of these components can be used to estimate the fluid velocity inside the shock wave. The observed dip is slightly bluer than the simulations, suggesting a faster shock velocity. A detailed parameter survey is still needed to produce a model with larger shock velocity without changing the light curve shape. On the other hand, it seems to be difficult for our model to reproduce the photon excess around 657 -- 660 nm in the observation, because this excess is thought to be composed of H$\alpha$ photons that are emitted in the CSM in front of the shock, traveling toward the shock, and redirected to the observer by electron scattering or H$\alpha$ absorption/re-emission with a positive line-of-sight velocity of several thousand km s$^{-1}$ in the shocked CSM. 
However, due to the high optical depth inside the shock wave, such a photon would immediately be randomly oriented due to electron scattering. Therefore, a non-spherically symmetric CSM (as suggested by \citealt{Leonard00} from spectropolarimetry) is probably needed to reproduce this red-side photon excess. 

SN 1997ab is one of the few SN IIn where high-resolution echelle spectra had been taken. Inspecting the spectra taken around 1 year after discovery \citep{salamanca1998}, the co-existence of the P-Cyg line overlaid with the broad emission line suggests an eruptive origin for the dense CSM generating the lines. However, a notable difference from our resultant spectra is that the peak of the broad line in SN 1997ab is shifted to the blue side relative to that of the narrow P-Cyg emission line, whereas both of the peak positions are coincident in our simulations. 
\cite{salamanca1998} explain that the red side of the broad emission line is reduced because H$\alpha$ photons are self-absorbed in the red side resulting in the blue-shifted broad emission line. 
Based on our simulations, the presence of the narrow P-Cyg line suggests that eruptive mass-loss occurs. Time from the start of eruption to the present is written as $t_{\rm{erup}} = t_{\rm{obs}} \times (v_{\rm{sh}}/v_{\rm{CSM}})$, with the observation time of $t_{\rm{obs}} \sim 1$ yr and $v_{\rm{CSM}} \sim 90$ km/s, which is estimated from the width of narrow P-Cyg line in \cite{salamanca1998}.  Although we cannot estimate the shock wave velocity, $v_{\rm{sh}}$, because the blue-shifted excess is not observed in SN1997ab, we obtain $t_{\rm{erup}} \sim$ several decades for $v_{\rm{sh}}$ of several thousand km/s. 
The spectral features of this SN will be studied in future work, with a refined modelling of the CSM that reproduces the luminosity evolution.

KISS15s is a long-lasting SN IIn with a clear blue-shifted excess in the spectra from day 124.2 to day 431.4 (their Figure 20), which is also seen in our model at late phase. We compare the low-wavelength edge of the observed excess with the expected maximal wavelength shift from our model (Equation \ref{eqn:lambda_max_scat}), and find $v_{\rm down}\approx 3500$ km s$^{-1}$. From the Rankine-Hugoniot relation under the strong shock limit, this translates to a shock velocity of $v_{\rm sh}\approx 4100$--$4700$ km s$^{-1}$ for an adiabatic index of $\gamma=4/3$--$5/3$. This is a factor of $\approx 2$ larger than $v_{\rm sh}\approx 2000$ km s$^{-1}$ adopted by \cite{kokubo2019} from the width of the intermediate-width lines. The change in $v_{\rm sh}$ drastically affects the inferred mass-loss rate and total shocked CSM mass within a duration $t_{\rm duration}$, based on equations (19) and (20) of \cite{kokubo2019},
\begin{eqnarray}
    \dot{M} &\sim& 0.4\ {\rm M_\odot\ yr^{-1}} \left(\frac{L_{\rm bol}}{0.8\times 10^{43}\ {\rm erg\ s^{-1}}}\right) \left(\frac{v_w}{40\ {\rm km\ s^{-1}}}\right)\left(\frac{\epsilon}{0.3}\right)^{-1}\left(\frac{v_{\rm sh}}{2000\ {\rm km \ s^{-1}}}\right)^{-3}\\
    M_{\rm CSM, 600d} &\sim & 35\ {\rm M_\odot}\left(\frac{L_{\rm bol}}{0.8\times 10^{43}\ {\rm erg\ s^{-1}}}\right) \left(\frac{t_{\rm duration}}{600\ {\rm days}}\right)\left(\frac{\epsilon}{0.3}\right)^{-1}\left(\frac{v_{\rm sh}}{2000\ {\rm km \ s^{-1}}}\right)^{-2},
\end{eqnarray}
where $L_{\rm bol}$ is the bolometric luminosity which is nearly constant ($0.8$--$0.9\times 10^{43}$ erg s$^{-1}$) from day 100 to 600, $v_w\lesssim 47.2$ km s$^{-1}$ is the CSM velocity constrained from high-resolution spectra, and $\epsilon$ is the radiation conversion efficiency. With other parameters fixed, the inferred $\dot{M}$ and $M_{\rm CSM, 600d}$ for our updated $v_{\rm sh}$ are drastically reduced to
\begin{eqnarray}
    \dot{M} &\sim& 0.04\ {\rm M_\odot\ yr^{-1}} \left(\frac{L_{\rm bol}}{0.8\times 10^{43}\ {\rm erg\ s^{-1}}}\right) \left(\frac{v_w}{40\ {\rm km\ s^{-1}}}\right)\left(\frac{\epsilon}{0.3}\right)^{-1}\left(\frac{v_{\rm sh}}{4400\ {\rm km \ s^{-1}}}\right)^{-3}\\
    M_{\rm CSM, 600d} &\sim& 7\ {\rm M_\odot}\left(\frac{L_{\rm bol}}{0.8\times 10^{43}\ {\rm erg\ s^{-1}}}\right) \left(\frac{t_{\rm duration}}{600\ {\rm days}}\right)\left(\frac{\epsilon}{0.3}\right)^{-1}\left(\frac{v_{\rm sh}}{4400\ {\rm km \ s^{-1}}}\right)^{-2} .
\end{eqnarray}
Albeit the possible uncertainty in $\epsilon$ the inferred CSM mass is found to be more compatible with the envelopes of RSGs, and thus may be a more natural explanation given by the low $v_w$ that may favor an RSG over an LBV. The inferred mass-loss rate indicates a Thomson scattering optical depth in the unshocked CSM of at most 
\begin{eqnarray}
    \tau_{\rm es} \approx \frac{\kappa \dot{M}}{4\pi v_w v_{\rm sh}t} \sim 2.2\left(\frac{\kappa_{\rm es}}{0.34\ {\rm cm^2\ g^{-1}}}\right) \left(\frac{L_{\rm bol}}{0.8\times 10^{43}\ {\rm erg\ s^{-1}}}\right) \left(\frac{\epsilon}{0.3}\right)^{-1}\left(\frac{v_{\rm sh}}{4400\ {\rm km \ s^{-1}}}\right)^{-4} \left(\frac{t}{200\ {\rm day}}\right)^{-1}
\end{eqnarray}
where $\kappa_{\rm es}$ is the Thomson scattering opacity. We note that this is an upper limit, under the assumption that the entire unshocked CSM is ionized. From equation (\ref{eq:dlambda}), we expect that scattering in this CSM creates an emission line with half-width $\sim 1000$ km s$^{-1}$, which is roughly consistent with the observed intermediate-width line.

\subsection{Caveats and Avenues for Future Work}
In this work we have done the Monte-Carlo simulations as post-process, assuming that the structure of the CSM does not change while the emitted Monte Carlo packets move in the computational region. We have tested the convergence of our results by changing the duration of sampling the packets by half shorter and an order longer, finding similar results. However, the post-process approach would not work when the hydrodynamical profiles of the shocked region and CSM significantly change during the diffusion time in the CSM, such as before the SN has reached its luminosity peak.
Future work will consider time-dependence of the hydrodynamical profiles of the CSM and the shocked region, which may help when modelling the very early phase of the SNe uncovered by e.g. flash spectroscopy \citep{Khazov_2016,yaron2017,Bruch_2021}.

The fractions of the ionization and excitation states are calculated assuming the local thermodynamic equilibrium with the gas temperature. This is a rough approximation in the CSM outside the photosphere. Thus we have discussed the information that can be deduced from the shape of the H$\alpha$ line by an analytical method based on results of the Monte Carlo simulations. As a consequence, we find that the velocity structure of the CSM can be inferred from the line shape.

There are claims that a significant fraction of SNe IIn show evidence for asymmetric CSM \citep[e.g.,][]{Leonard00,Bilinski18,Soumagnac20}. Our simulations using Monte-Carlo radiation transport can be straightforwardly extended to asymmetric CSM, by coupling to multi-dimensional hydrodynamical simulations of SNe interacting with aspherical CSM \citep{Vlasis16,McDowell18,Suzuki19}.

We have discussed H$\alpha$ spectra from a limited set of models, composed of a specific SN explosion with two CSM structures having a similar density distribution. It is known that SNe IIn have a diversity in their light curves and spectral evolution \citep[see e.g., ][for one such categorization]{taddia2013}, probably due to the diversity in the extent and mass of the CSM. While we adopted parameters of SN ejecta and CSM that lie close to the representative light curve properties of SN IIn, we did not attempt to fit our results with actual observations of SN IIn. In future work we plan to conduct a more comprehensive study on comparison of our model with observed SN IIn events, including the two events SN 1997ab and KISS-15s mentioned in this paper.

\begin{acknowledgements}
We thank Akihiro Suzuki for helpful comments throughout the advancement of this work. This work is supported by JSPS KAKENHI Grant Numbers 22K03688, 22K03671, 21J13957, 20H05639, 19H00693 MEXT, Japan, and by the Sasakawa Scientific Research Grant from The Japan Science Society. DT is supported by the Sherman Fairchild Postdoctoral Fellowship at the California Institute of Technology. 
Numerical computations were in part carried out on Cray XC50 at Center for Computational Astrophysics, National Astronomical Observatory of Japan.
\end{acknowledgements}

\software{ {MESA \citep{Paxton_2011,Paxton_2013,Paxton_2015,Paxton_2018,Paxton_2019,Jermyn23}, CHIPS \citep{takei2022,Takei_et_al_2023}}}

\appendix

\section{Calculation method of optical depth for H$\alpha$ absorption}
\label{app:tau_ha}

Here we summarize the procedure to calculate the optical depth for the H$\alpha$ transition using the Sobolev approximation. The cross section for H$\alpha$ absorption is \citep{rybicki1979}
\begin{eqnarray}
    \sigma_{\rm{H\alpha}} = \frac{\pi e^2}{m_e c} f_{\rm{23}} \phi(\nu), \nonumber
\end{eqnarray}
where $f_{\rm{23}}$ is the oscillator strength for H$\alpha$ and $\phi(\nu)$ is the line profile function of the photon frequency $\nu$. 
In the Sobolev approximation, the line profile function is replaced with the delta function as follows, 
\begin{eqnarray}
    \sigma_{\rm{H\alpha}} = \frac{\pi e^2}{m_e c} f_{\rm{23}} \delta(\nu_{\rm{0}} - \nu_{\rm{H\alpha}}). \nonumber
\end{eqnarray}
The subscript 0 indicates that the quantities are defined in the fluid comoving frame, and $\sigma_{\rm{H\alpha}}$ has a non-zero value when the frequency of a photon $\nu_{\rm{0}}$ in the fluid comoving frame is equal to $\nu_{\rm{H\alpha}}$. 
Using this, the optical depth for H$\alpha$ is as follows, 
\begin{eqnarray}
    \label{eq:tau0_int}
    \tau_{\rm{H\alpha}} &=& \int \sigma_{\rm{H\alpha}}\ n_{\rm{H\alpha}}\ d s \nonumber \\ 
&=&\int\sigma_{\rm H\alpha}n_{\rm H\alpha}\left(\frac{d\nu}{ds}\right)^{-1}d\nu,
\end{eqnarray}
where $d s$ is the infinitesimal path length of a photon. 
Thus, the optical depth becomes zero if  $\nu_{\rm H\alpha}=\nu_0=\nu\gamma(1-v\cos\theta/c)$ never holds on this path, where $\nu$ denotes the photon frequency in the observer frame and $\gamma$ denotes the Lorentz factor. Otherwise the optical depth can be obtained by performing the integral as
\begin{equation}
\label{eq:tauHalpha}
    \tau_{\rm{H\alpha}}= \frac{\pi e^2}{m_e c} f_{\rm{23}}n_{\rm H\alpha}\left(\frac{d\nu}{ds}\right)^{-1}.
\end{equation}
\begin{figure}
    \centering
    \includegraphics[scale=0.3]{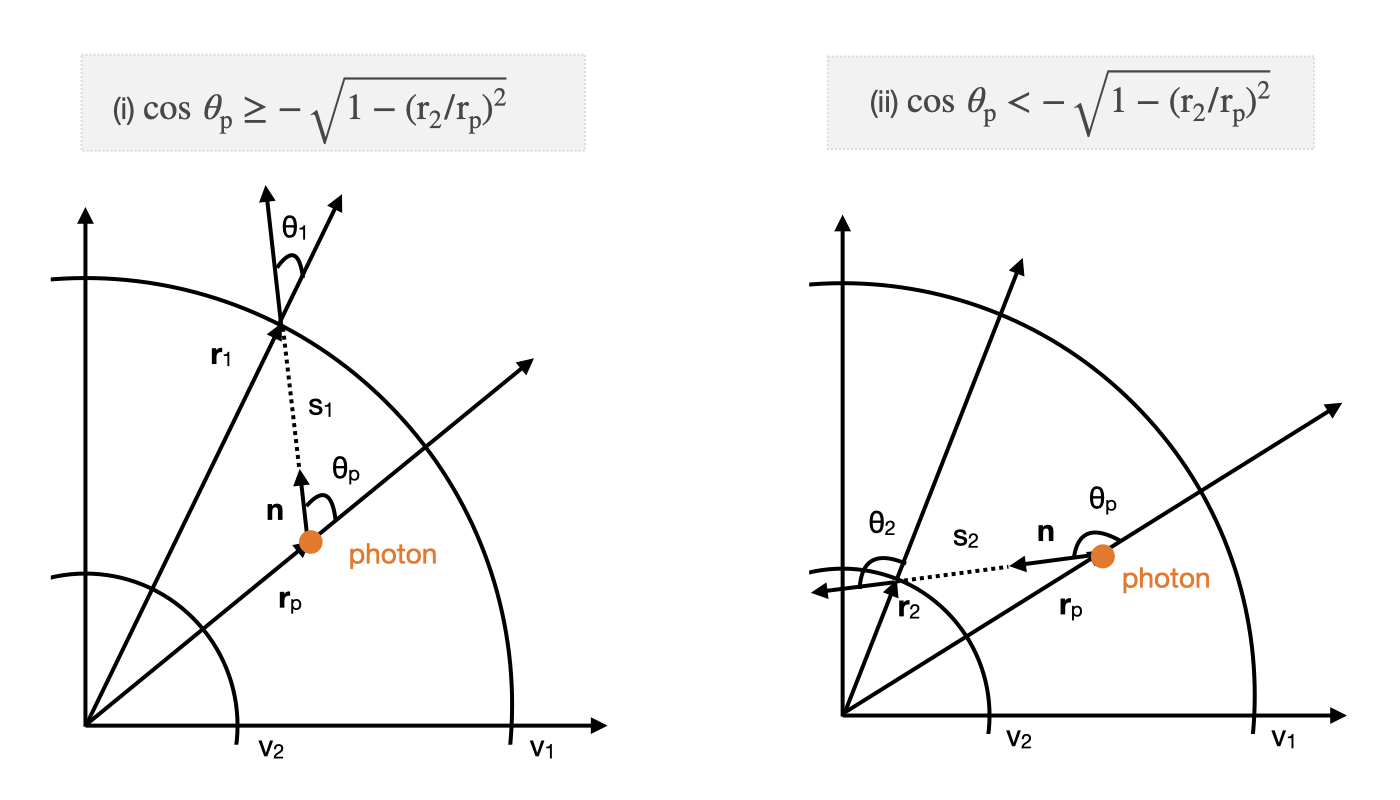}
    \caption{Schematic pictures of two cases of the photon transport.}
    \label{fig:tauha_sobolev}
\end{figure}
We evaluate $ds/d\nu$ by taking values of the frequency of the H$\alpha$ photon at the both edges of the path $\delta s$ from a point p to $i$ ($i=1,$ or 2. See Fig. \ref{fig:tauha_sobolev}) as
\begin{equation}
    \label{eq:dsdnu}
    \left(\frac{d\nu}{ds}\right)^{-1} \simeq \frac{\delta s}{\left|\nu_{{\rm{H\alpha,}}i}-\nu_{\rm{H\alpha,p}}\right|},
\end{equation}
where $\nu_{{\rm{H\alpha,}}i}=\nu_{H\alpha}/\gamma(r_{i})(1-v(r_i)\cos\theta_i/c)$ and $\nu_{\rm{H\alpha,p}}=\nu_{H\alpha}/\gamma(r_{\rm{p}})(1-v(r_{\rm p})\cos\theta_{\rm p}/c)$. 
Substituting this derivative into Equation (\ref{eq:tauHalpha}), we obtain 
\begin{equation}\label{eq:opticaldepth}
    \tau_{\rm{H\alpha}} = \frac{\pi e^2}{m_{\rm{e}} c} f_{\rm{23}}\ n_{\rm{H\alpha}}\ \frac{\delta s}{\left|\nu_{{\rm{H\alpha,}}i}-\nu_{\rm{H\alpha,p}}\right|}.
\end{equation}
 Suppose that a photon in a certain cell travels from a point p in the direction $\theta_{\rm p}$. The cell is bounded by two circles at $r=r_{\rm 2}$ and $r=r_{\rm 1}$ ($r_{\rm 1}>r_{\rm 2}$). We consider two different cases as shown in Figure \ref{fig:tauha_sobolev} to calculate the angle $\theta_i$ at the cell boundary.
\\

\noindent
(i) The case of cos $\theta_{\rm{p}}\geq  -\sqrt{1-(r_2/r_{\rm p})^2}$

The photon reaches the outer boundary of a cell by traveling a path $s_{\rm{1}}$. 
Let $v_{\rm{1}}$ and $v_{\rm{2}}$ denote the flow velocities at the outer and inner boundaries, respectively. 
The position vector of the photon is ${\bf r_{\rm{p}}} = \left[x_{\rm{p}}, y_{\rm{p}}, z_{\rm{p}}\right]$, the unit vector in the direction of the photon is ${\bf n} = \left[ n_{\rm{x}}, n_{\rm{y}}, n_{\rm{z}} \right]$, and the position vector at which the photon reaches the outer boundary is ${\bf r_{\rm{1}}} = \left[x_{\rm{1}}, y_{\rm{1}}, z_{\rm{1}}\right]$. 
Using the cosine formula, we obtain
\begin{eqnarray}
    | {\bf r_{\rm{1}}} |^2 = s_{\rm{1}}^2 + | {\bf r_{\rm{p}}} |^2 - 2 s_{\rm{1}} | {\bf r_{\rm{p}}} | \cos\ (\pi - \theta_{\rm p} ) \nonumber, 
\end{eqnarray}
where $r_1=|{\bf r_1}|$ and $r_{\rm p}=|{\bf r_{\rm p}}|$. Solving this for $s_{\rm{1}}$, we obtain 
\begin{eqnarray}
    \delta s=s_{\rm{1}} = - r_{\rm{p}}\ \cos\theta_{\rm p} + \sqrt{r_{\rm{p}}^2\ \cos^2\theta_{\rm p} - ( r_{\rm{p}}^2 -r_{\rm{1}}^2 ) }. \nonumber
\end{eqnarray}
Using $s_{\rm{1}}$, we rewrite ${\bf r_{\rm{1}}}$ as 
\begin{eqnarray}
    {\bf r_{\rm{1}}} = {\bf r_{\rm{p}}} + s_{\rm{1}} {\bf n} = 
    [ x_{\rm{p}}+s_{\rm{1}}n_{\rm{x}},\  y_{\rm{p}}+s_{\rm{1}}n_{\rm{y}},\  z_{\rm{p}}+s_{\rm{1}}n_{\rm{z}} ].  \nonumber
\end{eqnarray}
Therefore, 
\begin{eqnarray}
    \rm{cos}\ \theta_{\rm{1}} 
    &=& \frac{{\bf r_{\rm{1}} \cdot {\bf n}}}{r_{\rm{1}}} \nonumber \\
    &=& \frac{(x_{\rm{p}}+s_{\rm{1}}n_{\rm{x}}) n_{\rm{x}} + (y_{\rm{p}}+s_{\rm{1}}n_{\rm{y}}) n_{\rm{y}} + (z_{\rm{p}}+s_{\rm{1}}n_{\rm{z}})  n_{\rm{z}}}{r_{\rm{1}}}. \nonumber
\end{eqnarray}
Here, we interpolate the flow velocity between the values of $v_{\rm{1}}$ and $v_{\rm{2}}$ to obtain $v_{\rm{p}}$ at the current photon position. 
\\

\noindent
(ii) The case of cos $\theta_{\rm{p}} < -\sqrt{1-(r_2/r_{\rm p})^2}$

The photon reaches the inner boundary of a cell by traveling a path $s_{\rm{2}}$, which corresponds to $\delta s$. 
Let ${\bf r_{\rm{2}}} = \left[x_{\rm{2}}, y_{\rm{2}}, z_{\rm{2}}\right]$ be the position vector of the inner boundary where the photon reaches, we obtain from the cosine formula as in (i) 
\begin{eqnarray}
    | {\bf r_{\rm{2}}} |^2 = s_{\rm{2}}^2 + | {\bf r_{\rm{p}}} |^2 - 2 s_{\rm{2}} | {\bf r_{\rm{p}}} | \rm{cos}\ (\pi - \theta ) \nonumber.
\end{eqnarray}
Solving this for $s_{\rm{2}}$, we obtain 
\begin{equation}
    \label{eq:s2}
    \delta s=s_{\rm{2}} = - r_{\rm{p}}\ \rm{cos}\ \theta_{\rm{p}} - \sqrt{r_{\rm{p}}^2\ \rm{cos}^2\ \theta_{\rm{p}} - ( r_{\rm{p}}^2 -r_{\rm{2}}^2 ) }.
\end{equation}
Then we can rewrite ${\bf r}_2$ as 
\begin{eqnarray}
    {\bf r}_2 = {\bf r_{\rm{p}}} + s_{\rm{2}} {\bf n} = [ x_{\rm{p}}+s_{\rm{2}}n_{\rm{x}},\  y_{\rm{p}}+s_{\rm{2}}n_{\rm{y}},\  z_{\rm{p}}+s_{\rm{2}}n_{\rm{z}} ]. \nonumber
\end{eqnarray}
Therefore, the cosine of the angle $\theta_2$ between the photon's path and the radial direction (see Fig. \ref{fig:tauha_sobolev}) can be obtained as
\begin{eqnarray}
    \rm{cos}\ \theta_{\rm{2}} 
    &=& \frac{{\bf r_{\rm{2}} \cdot {\bf n}}}{r_{\rm{2}}} \nonumber \\
    &=& \frac{(x_{\rm{p}}+s_{\rm{2}}n_{\rm{x}}) n_{\rm{x}} + (y_{\rm{p}}+s_{\rm{2}}n_{\rm{y}}) n_{\rm{y}} + (z_{\rm{p}}+s_{\rm{2}}n_{\rm{z}})  n_{\rm{z}}}{r_{\rm{2}}}. \nonumber
\end{eqnarray}
Thus we calculate the optical depth $\tau_{\rm H\alpha}$ for the H$\alpha$ transition from equation (\ref{eq:opticaldepth}) by substituting $\delta s$ and the frequencies in each case.

\section{Test calculation of the P-Cyg line}
\label{app:jeffery}

\begin{figure}
    \centering
    \includegraphics[scale=1.0]{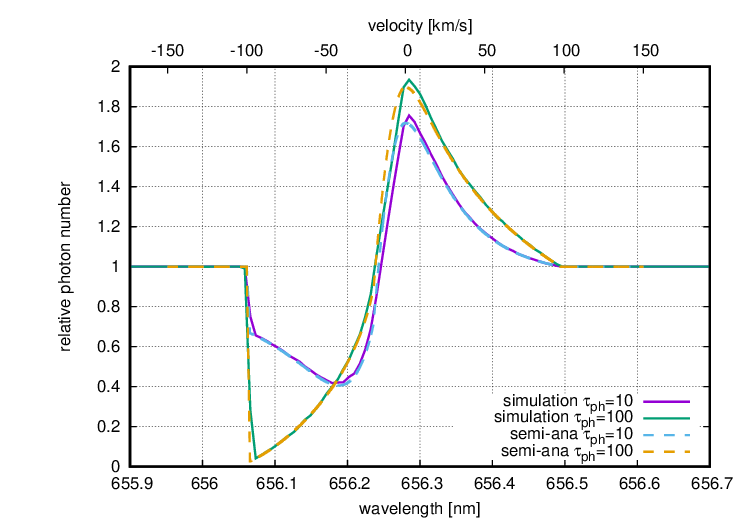}
    \caption{Comparison of P-Cyg lines between simulations and the semi-analytic calculations of \cite{jeffery1990}.}
    \label{fig:jeffery}
\end{figure}

In order to verify our Monte Carlo calculations, we performed test calculations to reproduce the P-Cyg line from a spherically symmetric flow-field undergoing homologous expansion. The simulated H$\alpha$ spectra are compared with those obtained from the semi-analytical model in \cite{jeffery1990}. 

The input of the \cite{jeffery1990} model is a homologously expanding medium, spanning from an arbitrarily defined ``photosphere" to the outer edge, which reprocesses the continuum emission from the photosphere. 
Here, we choose the parameters of this medium to be similar to the CSM that are relevant to this paper. Specifically, the radius of the photosphere is set to $10^{15}$ cm and that of the edge of the computational domain is set to $5 \times 10^{15}$ cm. The flow velocities at the photosphere and the outer edge are set to 20 km s$^{-1}$ and 100 km s$^{-1}$, respectively. 
Photons are injected from the photosphere with a blackbody spectrum, and are initially assigned to a frequency range $0.999 \nu_{\rm{H\alpha}} \leqq \nu_{\rm{0}} \leqq 1.002 \nu_{\rm{H\alpha}}$ in the fluid comoving frame.

The number density of hydrogen atoms that cause H$\alpha$ absorption is assumed to follow a power-law profile of $n_{\rm H\alpha}\propto r^{-2}$. The optical depth $\tau_{\rm{ph}}$ of H$\alpha$ absorption at the photosphere is defined in the semi-analytic model as 
\begin{equation}
    \tau_{\rm{ph}} = \frac{\pi e^2}{m_{\rm{e}} c} f n_{\rm{H\alpha,ph}} \lambda_{\rm{H\alpha}} \left| \frac{dv}{dr} \right|^{-1}. 
\end{equation}
We perform simulations with $\tau_{\rm{ph}}=10$ and $100$, under the assumption that each photon experiences at most only one H$\alpha$ absorption/re-emission during its transport. 
To compute a single spectrum we use $10^7$ photon packets, which are initially emitted from the photosphere and sampled at the outer edge of the computational domain. 

Figure \ref{fig:jeffery} compares P-Cyg lines  calculated by the Monte-Carlo simulations and the semi-analytic model. For both cases of $\tau_{\rm{ph}}=$ 10 and 100, the simulated spectra are in good agreement with the semi-analytical results. Our radiative transfer code reproduces the P-Cyg line, as well as results of previous works \citep{roth2015, wagle2023}. 

\bibliography{sample631}{}
\bibliographystyle{aasjournal}

\end{document}